\pdfoutput=0
\documentclass[apj]{emulateapj}
\usepackage{apjfonts}
\usepackage{xspace}
\usepackage{amsmath}
\usepackage{color}
\usepackage{multirow}
\usepackage{framed} 
\newcommand{\mtwom}{\ensuremath{M_{200m}}}
\newcommand{\mwl}{\ensuremath{M_{\mathrm{WL}}}}
\newcommand{\rfivec}{\ensuremath{r_{500c}}}
\newcommand{\rtwoc}{\ensuremath{r_{200c}}}
\newcommand{\rtwom}{\ensuremath{r_{200m}}}
\newcommand{\mfivec}{\ensuremath{M_{500c}}}
\newcommand{\mtwoc}{\ensuremath{M_{200c}}}

\newcommand{\kms}{km s\ensuremath{^{-1}}}
\newcommand{\hgpc}{\ensuremath{h^{-1}\mathrm{Gpc}}}
\newcommand{\hmpc}{\ensuremath{h^{-1}\mathrm{Mpc}}}
\newcommand{\hkpc}{\ensuremath{h^{-1}\mathrm{kpc}}}
\newcommand{\hmsun}{\ensuremath{h^{-1}\,{\it M}_{\odot}}}

\slugcomment{Accepted to the Astrophysical Journal}
\journalinfo{The Astrophysical Journal, in press}
\shorttitle{On the Accuracy of Weak Lensing Cluster Mass Reconstructions}
\shortauthors{Becker \& Kravtsov}

\begin{document}

\title{On the Accuracy of Weak Lensing Cluster Mass Reconstructions}

\author{Matthew R. Becker\altaffilmark{1,2}}
\author{Andrey V. Kravtsov\altaffilmark{2,3,4}}

\altaffiltext{1}{Department of Physics, 5720 S. Ellis Avenue, The University of Chicago, Chicago, IL 60637}
\altaffiltext{2}{Kavli Institute for Cosmological Physics, 5640 South Ellis Avenue, The University of Chicago, Chicago, IL 60637}
\altaffiltext{3}{Department of Astronomy and Astrophysics, 5640 South Ellis Avenue, The University of Chicago, Chicago, IL 60637}
\altaffiltext{4}{Enrico Fermi Institute, 5640 South Ellis Avenue, Chicago, IL 60637, USA}

\begin{abstract}
We study the bias and scatter in mass measurements of galaxy clusters
resulting from fitting a spherically-symmetric Navarro, Frenk \& White
model to the reduced tangential shear profile measured in weak lensing
observations. The reduced shear profiles are generated for
$\approx10^{4}$ cluster-sized halos formed in a $\Lambda$CDM
cosmological $N$-body simulation of a 1 \hgpc\ box. In agreement with
previous studies, we find that the scatter in the weak lensing masses
derived using this fitting method has irreducible contributions from
the triaxial shapes of cluster-sized halos and uncorrelated
large-scale matter projections along the line-of-sight. Additionally,
we find that correlated large-scale structure within several virial
radii of clusters contributes a smaller, but nevertheless significant,
amount to the scatter. The intrinsic scatter due to these physical
sources is $\approx 20\%$ for massive clusters, and can be as high as
$\approx 30\%$ for group-sized systems. For current, ground-based
observations, however, the total scatter should be dominated by shape
noise from the background galaxies used to measure the
shear. Importantly, we find that weak lensing mass measurements can
have a small, $\approx 5-10\%$, but non-negligible amount of
bias. Given that weak lensing measurements of cluster masses are a
powerful way to calibrate cluster mass-observable relations for
precision cosmological constraints, we strongly emphasize that a
robust calibration of the bias requires detailed simulations which
include more observational effects than we consider here.  Such a
calibration exercise needs to be carried out for each specific weak
lensing mass estimation method, as the details of the method determine
in part the expected scatter and bias. We present an iterative method for estimating mass \mfivec\ that can eliminate the bias for analyses of ground-based data.

\end{abstract}


\section{Introduction}\label{sec:intro}

The abundance of galaxy clusters as a function mass and redshift can
be used to investigate both the nature of dark energy and, potentially,
deviations of gravity from General Relativity. The power in this cosmological test arises from the sensitivity of the abundance of galaxy clusters 
to both the geometry of the universe and the
growth of structure \citet[e.g.,][see \citeauthor{voit2005} \citeyear{voit2005}
for a recent review]{holder2001,haiman2001}.  While in principle a comparison between the predicted and observed abundance of galaxy clusters  is straightforward, there are complications which will
need to be at least partially addressed through simulations.  These
complications include the completeness and purity of cluster finding
algorithms
\citep[e.g.,][]{reblinsky1999,white2002,carlstrom2002,cohn2007,rozo2007a,cohn2009,vikhlinin2009b} 
and bias and scatter in estimators of total cluster mass \citep[][]{lima2005,shaw2010}. The total mass is critical because it
is by far the most accurate theoretically predicted cluster property.  

Although in principle the total mass-observable relations can be
calibrated self-consistently within a given cluster
sample \citep[e.g.,][]{majumdar2003,majumdar2004,lima2005}, there is also
significant hope that masses measured using weak lensing (WL) observations
can be used to accurately calibrate such relations \citep[e.g.,][and references
therein]{hoekstra2007,mahdavi2008,zhang2008,vikhlinin2009b,zhang2010}. 
Indeed, WL measurements of masses have now been made for dozens of clusters \citep[e.g.,][]{dahle2006,bardeau2007,hoekstra2007,berge2008,okabe2010,abate2009} and this number is expected to increase to hundreds of clusters in the near future \citep{kubo2009}. However, for this hope to be realized
in practice, we need to know both the scatter and potential bias in
the WL mass measurements. The scatter will determine the number
of WL mass measurements required to calibrate normalization
and slope of scaling relations to a given accuracy. The bias will
determine the systematic uncertainty with which such a calibration can
be made.

In this work we investigate the scatter and bias in WL estimates of
cluster masses obtained from fitting the Navarro-Frenk-White \citep[NFW,][]{navarro1997} profile to the shear profile around
individual clusters. Previous studies in the literature have
identified two primary systematic effects in WL masses.  First, the
triaxial shapes of Cold Dark Matter (CDM) halos
\citep[e.g.,][]{warren1992,jing2002,bailin2005,kasun2005,allgood2006,shaw2006,bett2007,maccio2007,maccio2008}
can bias a spherically-symmetric model fit of the reduced tangential shear profile and lead to errors of
$\approx\pm30-50\%$ in the estimated mass \citep{clowe2004,oguri2005,corless2007,meneghetti2010a}.  The essential sense of this effect is that for halos whose major axes are aligned along the line-of-sight (LOS), the WL mass is overestimated, and for halos whose major axes are transverse to the LOS, the WL mass is underestimated.  

Second, correlated and uncorrelated large-scale structure (LSS) along
the LOS can cause positive bias of $\approx 30\%$ and scatter of
$\approx20\%$ in the recovered WL
masses \citep[][]{metzler2001,hoekstra2001,hoekstra2003,hoekstra2010a,deputter2005,marian2010,noh2010}.
The exact amount of bias in the WL masses due to correlated LSS
depends strongly on the method used to analyze the WL data
(compare \citeauthor{metzler2001} \citeyear{metzler2001}
with \citeauthor{marian2010} \citeyear{marian2010} and our results
below).  Additionally, the estimated amount of scatter in the WL
masses due to correlated LSS depends somewhat on how much LSS along
the LOS is included from the
simulations \citep[][]{metzler2001}.  \citet[][]{hoekstra2001,hoekstra2003} and \cite{hoekstra2010a}
found that uncorrelated LSS along the LOS does not bias the WL masses
but does add extra scatter of $\approx15-30\%$ depending on the
cluster's mass.  Additionally, uncorrelated LSS from random
projections along the LOS introduces correlated noise in the shear field of the
clusters \citep{hoekstra2001,hoekstra2003,hoekstra2010a,dodelson2004}.

Note that the projection of LSS and the effects of triaxiality are closely associated. Neighboring halos of similar mass are generally connected by a filament of matter with the fraction of halos connected
by filaments dropping as the distance between halos is increased
\citep{colberg2005}. The direction of the major axis of halos is
correlated with the direction to its massive neighbor and the filament
connecting the halos
\citep[e.g.,][]{splinter1997,onuora2000,faltenbacher2002,hopkins2005,bailin2005,kasun2005,basilakos2006,aragon-calvo2007,lee2007,hahn2007,zhang2009}. Furthermore, these alignments persist out to very large scales, with the correlation finally reaching zero only at $\approx100$ \hmpc\ \citep[][]{faltenbacher2002,hopkins2005}.  Therefore,
halos viewed along their major axes would also be more likely to
exhibit a larger amount of filamentary LSS projecting onto the halo's
field (see \citeauthor{noh2010} \citeyear{noh2010} for a similar study which 
demonstrated this effect explicitly for many cluster observables including WL masses).

In this work, we extend these previous studies by explicitly and
systematically considering the effects of halo shape, as well as
correlated, and uncorrelated LSS on the WL mass estimates.  We aim not
only to give estimates of bias and scatter in the WL masses as a
function of 3D mass, defined within a spherical radius enclosing a
given overdensity, but also to synthesize these previous results with
our own into a coherent picture of the sources of scatter and bias in
WL masses.  To this end, we use the entire population of halos in
large $\Lambda$CDM simulation to study the relationship between WL
masses and 3D masses statistically. Additionally, we systematically
study this relationship under different amounts of structure projected
along the LOS.  Our treatment is different than the previous works
mentioned above because we simultaneously consider a large number of
integration lengths in the range of 3-400 \hmpc, employ a commonly
used WL mass estimator to enable easy comparison to current
observational studies, use a statistical sample ($\sim10^{4}$) of
halos at multiple redshifts (regardless of their dynamical state or
environment), and avoid simulation box replications and random
rotations by using the results of \citet{hoekstra2001,hoekstra2003} to
extend our results to the full LOS integration length back to the weak
lensing source redshift.  Furthermore, as we will show below,
predicting the bias in WL mass estimates to better than $10\%$ is a
non-trivial task that will require detailed simulation studies.  In
this context, our study serves as an example of the kind of work that
will be needed to obtain percent-level accuracy from future WL mass
estimates.

Although in practice different methods can be used to estimate the total mass using the observed shear field \citep{king2001b,hoekstra2003,dodelson2004,maturi2005,corless2008,marian2010,oguri2010}, in
this study we adopt a specific model in which the density profiles of
clusters are assumed to be described by the NFW profile.  The prediction for the reduced tangential shear in the thin-lens approximation based on this profile \citep{bartelmann1996,wright2000} is then used to fit the reduced tangential shear profiles of the halos from the simulations.  This method is common \citep[e.g.,][]{clowe2001,hoekstra2002,clowe2002,bardeau2005,jee2005,clowe2006,dahle2006,kubo2007,paulin-henriksson2007,pedersen2007,okabe2010,abate2009,umetsu2009,kubo2009,hamana2009,holhjem2009,israel2010} and serves to illustrate our main points.  Our results as to the scatter of the estimated WL masses with respect to the 3D masses are specific to this
method. Other methods require
their own quantitative evaluation of the WL mass errors using simulations along similar lines.  However, our results do have some applicability to aperture densitometry \citep[also known as $\zeta$-statistics,][]{fahlman1994,kaiser1995} WL mass measurements. It is clear from the study of \citet{meneghetti2010a} that WL masses estimated from aperture densitometry and spherically-symmetric fits to the reduced tangential shear profile are quite correlated. Thus we can expect qualitatively similar conclusions about the sources of scatter and bias in WL mass estimated from these two methods.

Note also that we have made no attempt to identify an optimal method
to estimate the mass from the shear field
\citep[e.g.,][]{dodelson2004,maturi2005,corless2008,marian2010}, 
which could potentially decrease the scatter or bias.  In fact,
optimal, compensated aperture mass
filters \citep{kaiser1994,schneider1996} applied to shear fields in
the context of WL peak finding have been shown to be able to largely
eliminate the effects of uncorrelated LSS on the mass reconstructions
of individual
peaks \citep{maturi2005,marian2010}.  \citet{dodelson2004} suggested
using the correlations in the noise of the shear field due to random
LSS projections to help remove this kind of contamination
(see \citeauthor{oguri2010} \citeyear{oguri2010} for a recent
observational study which employs a similar
technique).  \citet{corless2008} and \citet{corless2009} used triaxial halo models to account for the orientation
of the clusters along the LOS when fitting for WL masses using the entire two-dimensional shear field. 
They tested this procedure with analytic models for triaxial NFW halos and found that it can reduce the amount 
of bias in the WL masses through the use of a prior on the distribution of triaxial 
halo shapes from N-body simulations. Our results for the scatter and bias in the presence of observational errors,
triaxial halo shapes, and LSS may thus be somewhat pessimistic.  However,
given how commonly the mass measurement method we employ is used to
analyze observations, it is still necessary to obtain estimates of
scatter and bias in the relation between WL masses measured using this
method and three-dimensional masses.

In \S\ref{sec:siminfo} we describe the simulation and halo finder used
in this work.  In \S\ref{sec:lensmeth} we review the weak lensing
formalism and describe our procedure for extracting reduced tangential
shear profiles from the particle distributions around the halos in our
simulations.  In \S\ref{sec:results} we fit the reduced tangential
shear profiles of simulated clusters, illustrate how various sources
of scatter in the WL masses behave in simulations, and give estimates
of bias and scatter in the recovered masses in the presence of
observational errors.  In \S\ref{sec:discuss} we compare our results
to previous work and describe some implications of our bias and
scatter estimates.  Finally, we give some general remarks and
conclusions in \S\ref{sec:conc}.

\section{The cosmological simulation}\label{sec:siminfo}

For our study we use a simulation of a flat $\Lambda$CDM cosmology
with parameters consistent with the WMAP 7 year
results \citep{komatsu2010}: $\Omega_{m}=0.27$, $\Omega_{b}=0.044$,
$\sigma_{8}=0.79$, $h=0.7$ in units of 100
\kms$/$Mpc, and a spectral index of $n=0.95$.  The simulation followed
evolution of $1024^3$ dark matter particles from $z_i=60$ to $z=0$ in
a box of 1000 \hmpc\ on a side using distributed version of the
Adaptive Refinement Tree (ART
code \citet{kravtsov1997,gottloeber08}. The mass of each dark matter
particle in the simulation is $6.98\times10^{10}$ \hmsun\ and their
evolution was integrated with effective spatial resolution of
30 \hkpc. The simulation was used in \citet{tinker2008}, where it was
labeled L1000W.  We use the redshift 0.25 and 0.50 snapshots for our
results below.

The halos are identified using the spherical overdensity algorithm
described in \citet{tinker2008} and we refer the reader to this work for a complete description of the details concerning halo identification.
We measure the 3D halo mass using the common overdensity criterion
\begin{equation}\label{eqn:ovdef}
M_{\Delta}=\Delta\rho(z)\frac{4}{3}\pi r_{\Delta}^{3}
\end{equation}
where $M_{\Delta}$ is the mass at the overdensity $\Delta\rho(z)$ and
$r_{\Delta}$ is the radius enclosing this overdensity.  We use masses
defined using overdensities defined with respect to both the mean,
$\rho_{m}(z)$, and critical densities, $\rho_{c}(z)$, at a given
redshift of our simulation snapshot.  We will follow the notation
that \mfivec\ is the mass with $\Delta\rho(z)=500\rho_{c}(z)$, \mtwom\
is the mass with $\Delta\rho(z)=200\rho_{m}(z)$, etc.  The various
mass thresholds used in our analysis will be listed when they are
relevant.

For every halo we additionally fit the spherically averaged
three-dimensional density profile with the NFW profile.  We first bin
the profiles using the following procedure.  We sort the halo's
particles by distance to the halo center into ascending order.  We use
the halo center output from the halo finder.  We then group the
particles in bins so that there are at least 30 particles per bin
moving out in radius.  The radius of the bin is set to the average radius
of the particles in the bin.  The error in the density is computed
assuming Poisson statistics.  The density profiles are then fit with
an NFW profile using a $\chi^{2}$-fitting metric using the non-linear
least-squares Levenberg-Marquardt algorithm \citep{press1992}.

We also measure the halo's shape from the inertia tensor.  The inertia tensor is defined as \citep[e.g.,][]{shaw2006}
\begin{equation}\label{eqn:moi}
{\cal M}_{ij}=\frac{1}{N_{\Delta}}\sum_{n} x^{(n)}_{i}x^{(n)}_{j}\ ,
\end{equation}
where the sum extends over all $N_{\Delta}$ particles of the halo
within a predefined radius $r_{\Delta}$ and $x^{(n)}_{i}$ is the $i$th
coordinate ($i=1,2,3$) of the $n$th particle relative to the halo
center.  We diagonalize ${\cal M}_{i,j}$ and compute its
eigenvalues and eigenvectors.  The eigenvalues and eigenvectors are
sorted into ascending order.  The square roots of the eigenvalues will
be denoted by $\{a,b,c\}$ and we adopt the convention that $a<b<c$.
With our conventions the intermediate-to-major axis ratio is $b/c$ and
the minor-to-major axis ratio is $a/c$.  Finally, we define a
parameter from the axis ratios called $S$,
\begin{equation}\label{eqn:S}
S=\frac{a}{c}\ .
\end{equation}
For a perfectly spherical halo, $S=1$ and for a perfectly prolate halo, $S=0$.  This parameter does not distinguish between oblate and prolate halos, but halos in $\Lambda$CDM cosmologies are known to be preferentially prolate  \citep[e.g.,][]{shaw2006}.

Note that we do not iteratively measure the
triaxial axes, as is customary done. We also do not apply radial
weighting and do not remove subhalos, as is often done in more
sophisticated algorithms measuring halo
triaxiality \citep[e.g.,][]{bett2007,lau2010}. In this study we
are only interested in the general direction of each halo major axis,
and only use the measured axis ratio $S$ to rank-order halos by their
triaxiality (i.e. we do not use its absolute value). Thus our
simplified method for estimating axis ratios should be adequate for
our purposes.

\section{Lensing formalism}\label{sec:lensmeth}

The weak lensing equations follow from the linearized geodesic and
Einstein equations set in an homogeneous, isotropic, expanding
universe, with weak perturbations to the metric (see \citeauthor{dodelson2003} \citeyear{dodelson2003} for a pedagogical introduction).  We use the
approximation given by eqs.~7-9 in \citet{jain2000} in which 
the line-of-sight component of the Laplacian of the potential 
is neglected and straight-line photon paths are assumed.  Under this approximation and assuming a flat geometry, the
convergence can be calculated as \citep{metzler2001}
\begin{equation}\label{eqn:born}
\kappa = \frac{3}{2}\left(\frac{H_{o}\chi}{c}\right)^{2}\Omega_{m}\int_{0}^{1}{t(1-t)\frac{\delta}{a}}dt\ ,
\end{equation}
where $\chi$ is the comoving distance to the source, $\chi'$ is the
radial component of the photon's comoving position along its
unperturbed path, $t\equiv\chi'/\chi$, $\delta$ is the mass
overdensity, $a$ is the scale factor normalized to unity today,
$H_{o}$ is the Hubble constant, $c$ is the speed of light, and
$\Omega_{m}$ is the total matter density at $z=0$ in units of the present day critical density.  This equation is
commonly referred to as the Born approximation in the literature and
is applicable for sources at a single redshift. In the case of multiple
sources at different redshifts, the integral can simply be averaged over the normalized source redshift distribution.
Equation~\ref{eqn:born} highlights the dimensionless lensing kernel
$g(t)=t(1-t)$.  In general, this kernel function is broad in redshift
and weak lensing measurements of individual clusters can therefore be
easily affected by structures projected along the LOS.

Although more complicated ray-tracing schemes exist to evaluate the
convergence and shear along the actual curved photon paths
\citep[e.g.,][]{jain2000,vale2003,hilbert2009}, we choose to use the
Born approximation to simplify and speed up our calculations. Generally, 
one expects that the Born approximation will fail in high-convergence regions \citep[see e.g.,][]{vale2003}.
We use a ray-tracing code similar to that of \citet{hilbert2009} to check 
the accuracy of the Born approximation around our halos. Over the radial range of $1\arcmin$ 
to $25\arcmin$ at both $z=0.25$ and $z=0.50$ 
the Born approximation is accurate to $\lesssim1\%$ and so does not
compromise the accuracy of our WL masses. 

Working in the flat-sky approximation, the two components of the shear and the
convergence are related through derivatives of the lensing potential
$\psi$:
\begin{equation}\label{eqn:gamma1}
\gamma_{1}=\frac{1}{2}\left(\partial_{1}^{2}\psi-\partial_{2}^{2}\psi\right)\ ,
\end{equation}
\begin{equation}\label{eqn:gamma2}
\gamma_{2}=\partial_{12}^{2}\psi\ ,
\end{equation}
 \begin{equation}\label{eqn:poissonconv}
 \kappa=\frac{1}{2}\nabla_{\perp}^{2}\psi\ .
\end{equation}
Note that equation~\ref{eqn:poissonconv} is a two-dimensional Poisson equation for $\psi$ sourced by $2\kappa$.  Assuming vacuum boundary conditions, $\psi$ can be written as a convolution of the two-dimensional Green's function with the effective source $2\kappa$,
\begin{equation}\label{eqn:greens}
\psi = \frac{1}{\pi}\int d^{2}\mathbf{x}' \kappa(\mathbf{x}') \ln{|\mathbf{x}-\mathbf{x}'|}\ .
\end{equation}
The shear components defined in Equations \ref{eqn:gamma1} and \ref{eqn:gamma2} can be obtained from $\kappa$ by taking the appropriate derivatives of Equation~(\ref{eqn:greens}) with respect to the components of $\mathbf{x}$.  We choose to directly convolve the resulting kernels for $\gamma_{1}$,
\begin{displaymath}
\frac{1}{2\pi}\left(\partial^{2}_{1}\ln|\mathbf{x}-\mathbf{x}'|-\partial^{2}_{2}\ln|\mathbf{x}-\mathbf{x}'|\right) = \frac{(x_{2} - x_{2}')^{2} - (x_{1} - x_{1}')^{2}}{2\pi\left[(x_{1} - x_{1}')^{2} + (x_{2} - x_{2}')^{2}\right]^{2}}\ ,
\end{displaymath}
and for $\gamma_{2}$,
\begin{displaymath}
\frac{1}{\pi}\partial^{2}_{12}\ln|\mathbf{x}-\mathbf{x}'| = -\frac{(x_{1} - x_{1}')(x_{2} - x_{2}')}{\pi\left[(x_{1} - x_{1}')^{2} + (x_{2} - x_{2}')^{2}\right]^{2}}\ ,
\end{displaymath}
into the convergence field $\kappa$ using an FFT and
zero padding.

\subsection{Analysis of the Simulation}

We use Equation~\ref{eqn:born} to produce convergence maps around
clusters extracted from the simulation.  Using a single simulation
snapshot, we extract all of the particles around each cluster in a
$20\times20\times400$ \hmpc\ box. We use comoving distances in this work.  The $z$-axis direction is used for
the long axis of the box. To explore the effects of projected
correlated LSS, we vary the length of the long axis of the box from
3 to 400 \hmpc, but always keep the transverse size of the
analysis volume fixed.

For a given choice of the LOS length, we sum over all particles from
the simulation along the long dimension of the volume in order to
compute the integral in Equation~\ref{eqn:born}, accounting for the
periodic boundary conditions.  In the transverse directions, we use
the triangular-shaped cloud interpolation onto a projected 2D grid
which has a dimension of $512\times512$ cells.  The angle of each
particle from the cluster center is computed assuming the cluster
center is at the comoving distance corresponding to the simulation
snapshot redshift.  The source redshift $z_{s}$ is fixed at 1.0 in
this work.  We then compute the shear field from the convergence maps
according to the procedure described above.  We have 
checked that with a grid of $1024\times1024$ cells and a transverse box width of 
15 \hmpc\ (a factor of $\approx2.7\times$ better resolution) that our results for the bias
in the WL masses we find below are unchanged by $\lesssim1\%$.  The results for the scatter
and the slope of the $\mwl-M_{\Delta}$ relation are unchanged to this accuracy as well.

For every grid cell, we calculate the component of the shear
tangential to the radius vector connecting the cell under
consideration to the center of the halo.  The shear transforms as a
second-rank tensor under rotations, so that we can define
\begin{displaymath}
\gamma_{E}\equiv-\gamma_{1}\cos(2\theta)-\gamma_{2}\sin(2\theta)
\end{displaymath}
\begin{displaymath}
\gamma_{B}\equiv\gamma_{1}\sin(2\theta)-\gamma_{2}\cos(2\theta)\ ,
\end{displaymath}
where $\theta$ is the angle counter-clockwise from the positive
1-axis.  We have used the common $E$- and $B$-mode decomposition applied
around the halo center and the grid cell under consideration in the equations above, so that the tangential
shear satisfies $\gamma_{T}=\gamma_{E}$.  Under
the assumption of small distortions to galaxy shapes due to gravitational lensing, the average shape over a set of
galaxies will give a measurement of the reduced shear
$g_{1,2}=\gamma_{1,2}/(1-\kappa)$ (see e.g., Appendix A of
\citeauthor{mandelbaum2006} \citeyear{mandelbaum2006} for this result
and higher order corrections). Thus using the convergence, we compute the reduced tangential shear field, $g=\gamma_{T}/(1-\kappa)$, for use in our WL mass modeling, details of which we describe in the next subsection.

\subsection{The Noise Properties of the Shear Profile and Fitting Methods}\label{sec:fitmeth}

In order to accurately predict the scatter in the cluster
masses estimated from reduced tangential shear profile fitting, we must ensure
that our calculated profiles have similar noise properties to observed
profiles.  \citet{hoekstra2001,hoekstra2003} has established that the
noise in the reduced tangential shear profiles can be broken into two
components.  The first component is random noise due to the intrinsic 
noise in the shapes of the galaxies themselves and should
decrease with the number of galaxies used in each radial bin as
$\propto 1/\sqrt{N}$.  The second source of noise is due to
LSS (and also due to triaxiality of halos). However, 
\citet{hoekstra2001,hoekstra2003} only considered noise from 
LSS uncorrelated with the target lens.  This source of noise
does not decrease as the source galaxy density increases, and in general 
LSS introduces correlated noise in the shear field around a halo \citep{dodelson2004}. 
We neglect all other potential sources of noise or
systematic errors such as contamination of the
source galaxies with cluster members \citep[e.g.,][]{okabe2010}, and
photometric redshift errors \citep[e.g.,][]{mandelbaum2008c} or unknown source redshifts \citep[e.g.,][]{okabe2010}.  
The contamination of the source galaxies with cluster 
members can produce an $\approx-10\%$ bias in typical WL masses \citep{okabe2010}, 
though the exact magnitude of this effect depends sensitively on the contamination rate.  
Unknown source redshifts can introduce systematic biases in WL masses of $\approx\pm5-10\%$ 
as the source redshift is changed by $\pm0.2$ \citep{okabe2010}.  The improper use of photometric 
redshifts can introduce large $\approx\pm5-15\%$ biases in the estimated lensing 
critical surface density \citep{mandelbaum2008c} and thus WL masses.

For halos extracted from a cosmological simulation, the noise
and correlations due to LSS are already included.  We thus simply add
to the mean reduced tangential shear value of each radial bin the Gaussian noise due to the limited number of
background sources.  This noise has a zero mean and variance
\begin{equation}\label{eqn:staterr}
\sigma^2_{\rm s}=\frac{\sigma_{\rm e}^{2}}{n_{\rm gal}A}\ ,
\end{equation} 
where $\sigma_{\rm e}$ is the intrinsic shape noise of the sources,
$n_{\rm gal}$ is the surface density of source galaxies on the
sky, and $A$ is the area of the annulus.  Note that magnification and size bias will introduce changes in 
the effective number density of sources as a function of radius relative to the geometric 
expectation given above in Equation~\ref{eqn:staterr} \citep{schmidt2009a,schmidt2009b,schmidt2010,rozo2010}.  
For simplicity we will neglect this effect in this work.

We adopt $\sigma_{\rm e}=0.3$ for the intrinsic shape noise.  This value is typical for ground-based observations like those in \citet{okabe2010}.
Note that $\sigma_{\rm e}$ is the shape noise in the reduced tangential shear per galaxy.  
This amount of shape noise in the reduced tangential shear is roughly equivalent to a 
shape noise of 0.4 per shear component (i.e. $\sigma_{\rm e}\approx0.4/\sqrt{2}$).  We will use the following 
representative values for the source galaxy density $n_{\rm gal}$: $n_{\rm gal}=10$ galaxies/arcmin$^{2}$ for the Dark Energy
Survey\footnote{\url{http://www.darkenergysurvey.org/}} (DES) or similar 
observations \citep[e.g.,][]{hoekstra2008,okabe2010}, $n_{\rm gal}=20$ galaxies/arcmin$^{2}$ for
deep ground-based observations, and $n_{\rm gal}=40$ galaxies/arcmin$^{2}$ for very deep ground-based 
observations like the Large Synoptic Survey Telescope\footnote{\url{http://www.lsst.org/lsst}} (LSST) or 
space-based observations like those from the {\it Hubble Space Telescope} \citep[e.g.,][]{hoekstra2002} or Euclid\footnote{\url{ http://sci.esa.int/euclid}}.  
We also present results with no observational errors added to the reduced shear profiles in order to illustrate the
intrinsic scatter and bias in the WL masses at fixed 3D mass.

As stated in \S\ref{sec:intro}, we will estimate the 3D mass of the cluster with WL 
by fitting the tangential component of the reduced shear
profile with that predicted from the NFW profile in the thin lens
approximation.  In the fits we vary both the total mass and
concentration independently.  We use logarithmic binning in
radius. Below we will systematically test the effects of variations in
the number of bins used and the maximum radius of the fits.  For simplicity, we fix the minimum 
radius used for fitting the binned reduced tangential shear profiles to 1\arcmin\ at all redshifts.
This value is similar to that used in typical ground-based WL analyses \citep[see e.g.,][]{okabe2010}.  
Note however that for space-based observations the minimum fit radius can be as 
small as 0.5\arcmin\ \citep[see e.g.,][]{hoekstra2010}.  Tests with the higher resolution $1024\times1024$ cell grids 
indicate that the bias in the WL masses we find below decreases by $\approx1-2\%$ using a 0.5\arcmin\ 
inner fitting radius, indicating that the exact choice of inner fitting radius has a relatively small effect on our results.

We use a $\chi^{2}$-fitting metric and the non-linear least-squares
Levenberg-Marquardt algorithm throughout \citep{press1992}.  For the
comparison of our results to WL observations, a $\chi^{2}$-fit is
appropriate.  Specifically, the $\chi^{2}$-fitting metric is
\begin{equation}
\chi^{2}=\sum_{i=1}^{N}\left[\frac{g_{i}-g_{NFW}(r_{i},M_{\Delta},c_{\Delta})}{\sigma_{s}(r_{i})}\right]^{2}\ ,
\end{equation}
where $g_{i}$ is the reduced tangential shear averaged over the annulus at radius $r_{i}$, $g_{NFW}(r,M_{\Delta},c_{\Delta})$ is the prediction for the reduced tangential shear from an NFW profile at radius $r$ for mass $M_{\Delta}$ and concentration $c_{\Delta}$, and $\sigma_{s}(r)$ is the intrinsic shape noise given by Equation~\ref{eqn:staterr} for the sources in the bin at radius $r$.  The radius of each bin $r_{i}$ is computed from the average radius of the sources in each bin.  With this definition of the fitting metric, each radial bin is treated independently, although LSS will correlate the radial bins as discussed above.  We also neglect the contribution of $g$ to the overall shape noise of the sources.  Finally, we have verified that an unweighted $\chi^{2}$-fit,
\begin{displaymath}
\chi^{2}_{\rm uw}=\sum_{i=1}^{N}\left[g_{i}-g_{NFW}(r_{i},M_{\Delta},c_{\Delta})\right]^{2}\ ,
\end{displaymath}
or a fit that minimizes the summed absolute deviations from the model profile,
\begin{displaymath}
\chi^{2}_{\rm abs}=\sum_{i=1}^{N}\left|g_{i}-g_{NFW}(r_{i},M_{\Delta},c_{\Delta})\right|\ ,
\end{displaymath}
both produce similar or larger scatter in the WL masses at fixed 3D
mass.  However the fractional bias in the WL masses
varies by a few percent depending
on the choice of fitting metric as described below.

In order to assess the importance of shape noise in the WL mass errors, we need to compare the properties WL mass estimates with and without shape noise.  However, as 
stated above, the bias in the WL masses depends on relative weights of radial bins.  Thus, when estimating 
the intrinsic scatter and bias in the WL mass estimates (i.e. with no shape noise), we aim to preserve the relative weights of the 
radial bins by using the same $\chi^{2}$-fit weighted by the
observational errors in the denominator, but do not add observational noise to the mean reduced shear value 
of each bin in the numerator.  This procedure eliminates spurious differences in the scatter and bias of the WL 
masses due to the choice of fitting metric when comparing results with or without shape noise.

\begin{figure*}[t]
\begin{center}
\includegraphics[scale=0.4]{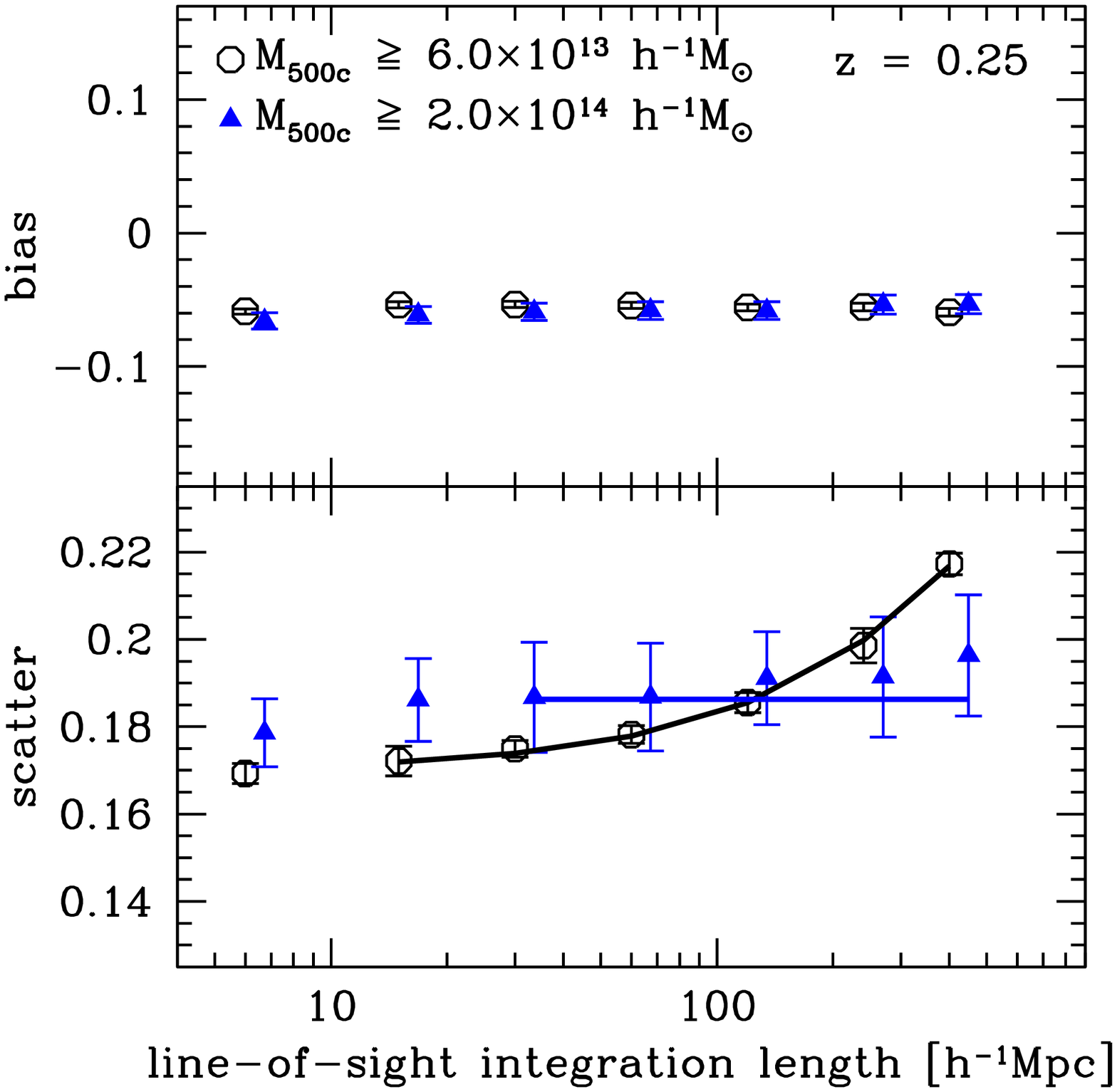}
\includegraphics[scale=0.4]{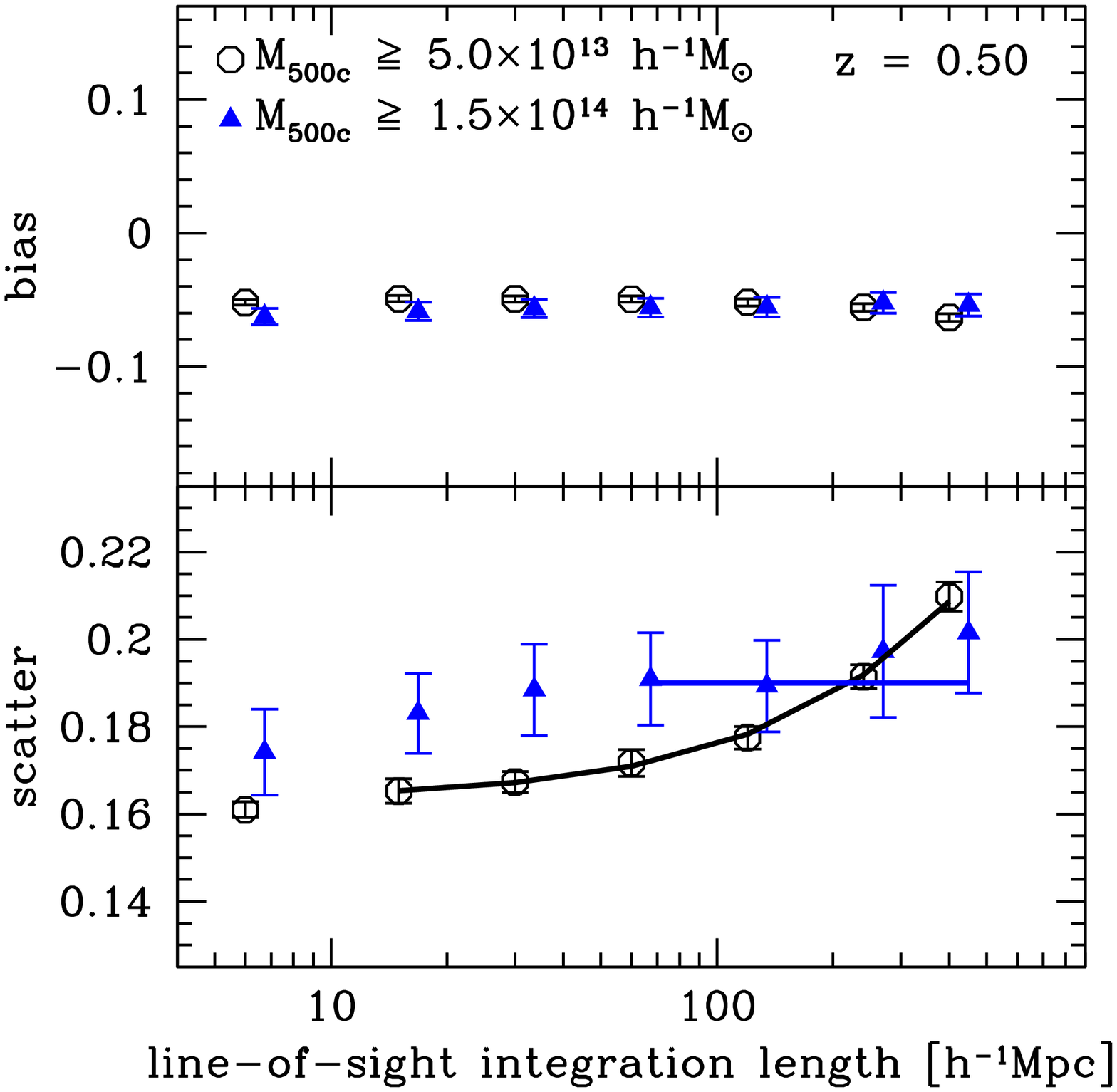}
\end{center}
\figcaption[]{The bias and scatter in WL mass measurements of \mfivec\ at $z=0.25$ (\textit{left}) and $z=0.50$ (\textit{right}). The open circles (\textit{black}) and filled triangles (\textit{blue}) show the results for halos satisfying the mass cuts given in the top panels for each redshift.  The filled triangles are shifted by $+0.05$ dex in the horizontal direction for clarity.  The lines show the predictions from Equation~\ref{eqn:fracscatmodel} in the regime where uncorrelated LSS along the LOS begins to have an effect on the scatter in the WL masses for each mass cut and redshift.  See \S\ref{sec:lssscat} for details concerning the fit of Equation~\ref{eqn:fracscatmodel} to the simulation data. The WL masses are clearly biased using for the NFW fitting and the behavior of the scatter as a function of LOS integration length is different for low- and high-mass halos. Note that a non-negligible fraction of the scatter and bias is due to correlated structures in matter distribution at distances between $\approx 6$ and $10$\hmpc\ from the cluster. \label{fig:scatdist}}
\end{figure*}

We compute the errors and correlations of various quantities derived
from the simulation and reduced tangential shear profile fits (e.g., the
parameters of and scatter in the \mwl-\mfivec\ relation) using a
jackknife method.  Namely, for the entire 1000 \hmpc\ simulation box we 
split the two dimensions perpendicular to the LOS
direction (i.e. the x-y plane perpendicular to the $z$-axis direction in our case) 
into a $10\times10$ grid of one hundred $100\times100$ \hmpc\ cells. We 
can then use these cells to compute jackknife estimates of the covariance 
matrix of our desired quantities by not considering the halos in each cell 
in turn (see \citeauthor{scranton2002} \citeyear{scranton2002} for a similar technique).

\section{Results}\label{sec:results}

\begin{figure*}[t]
\begin{center}
\includegraphics[scale=0.4]{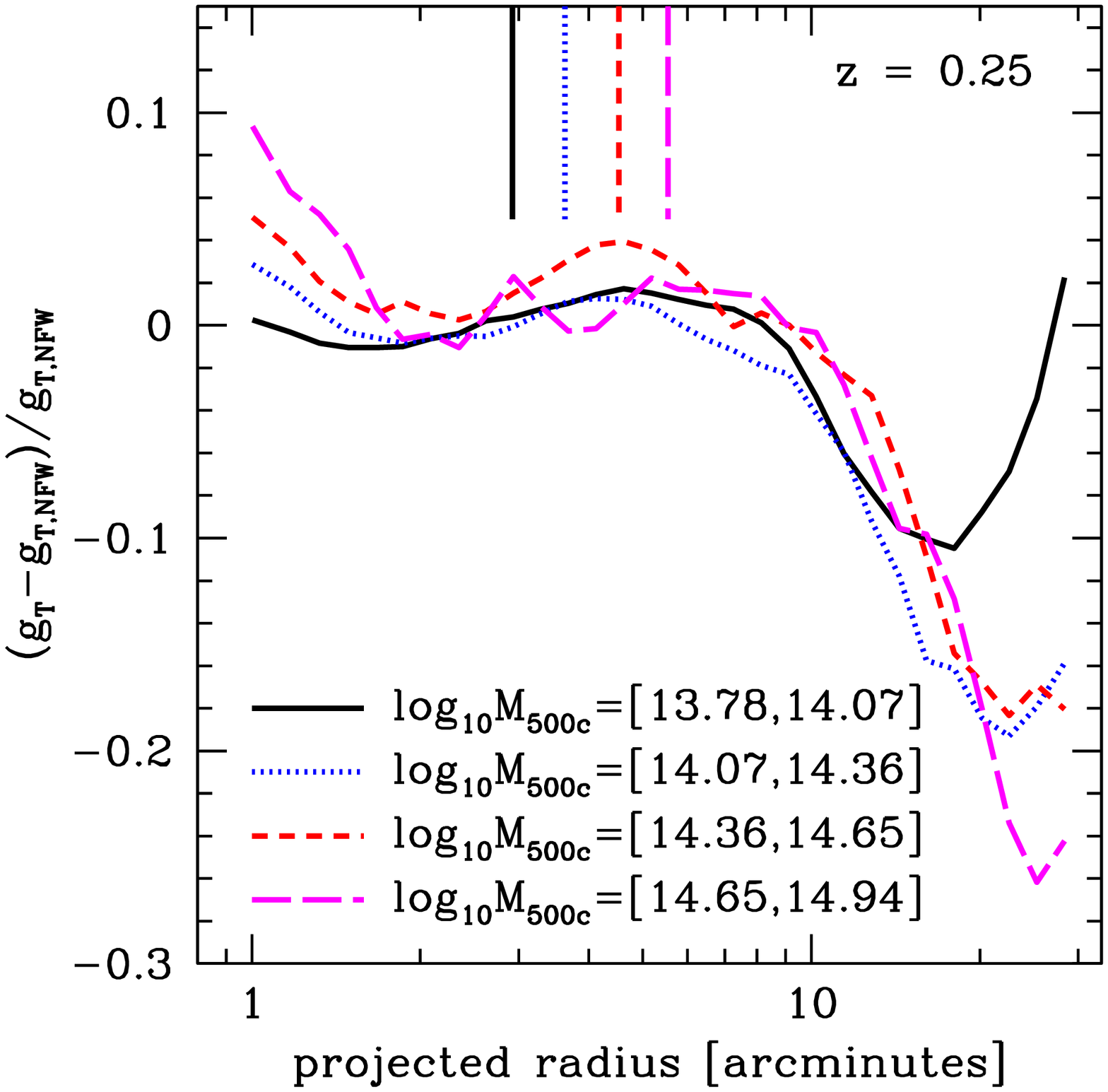}
\includegraphics[scale=0.4]{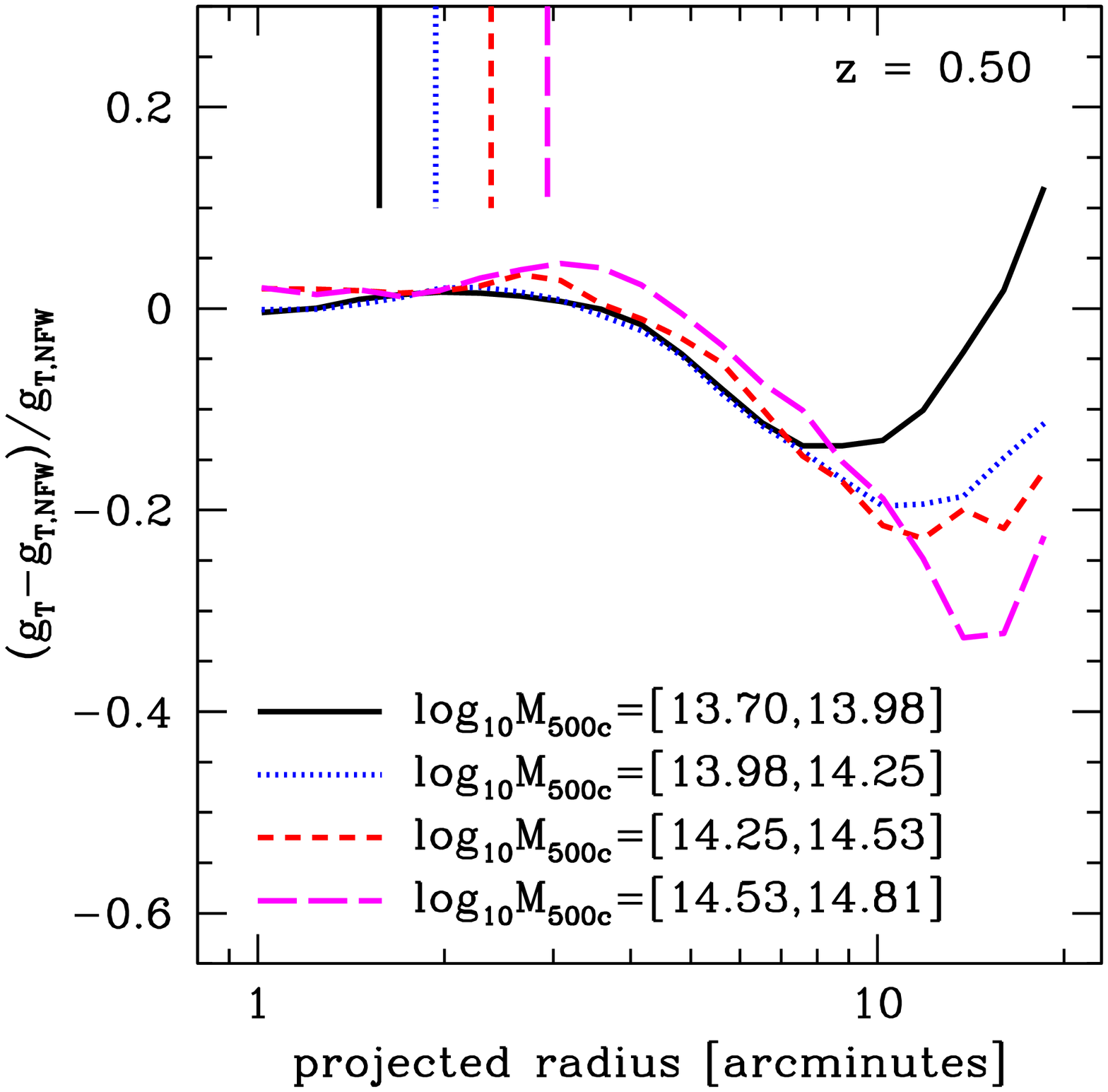}
\end{center}
\figcaption[]{Mean fractional difference between the reduced tangential shear profile predicted from an NFW fit to the three-dimensional mass profile of each halo, $g_{\rm T,NFW}$, and the halo's true reduced tangential shear profile, $g_{\rm T}$.  The average is taken in four bins of $\log\mfivec$ for the $z=0.25$ (\textit{left}) and $z=0.50$ (\textit{right}) snapshots.  The vertical lines extending from the top of each panel mark $r_{500c}$ for each halo mass bin.  The range in $\log\mfivec$ for each halo mass bin and the associated line style (and color) are given in the lower left corner of each panel.  The virial radius of the halos in each mass bin is at $\approx 2r_{500c}$. The up turn at large radii is due to 2-halo contributions to the stacked shear signal.  In the region between the 1- and 2-halo contributions beyond the halo virial radius, the NFW model is clearly wrong \protect\citep[see e.g.,][]{tavio2008}. Note that 1 arcminute corresponds to comoving distances of 206 \hkpc\ and 389 \hkpc\ at $z=0.25$ and $z=0.50$ respectively. \label{fig:gbias}}
\end{figure*}

Our aim in this section is to illustrate the effect of the
uncorrelated and correlated LSS on WL mass estimates and to give
estimates for the scatter in WL-measured masses at fixed 3D mass under a
variety of observational conditions.  Our primary results in
Figures~\ref{fig:scatdist}--\ref{fig:ensmass} are discussed in detail
in \S\ref{sec:lssscat} and concern the intrinsic scatter and bias in WL mass estimates 
in the absence of galaxy shape noise.  For these results we have chosen halos from
the $z=0.25$ and $z=0.50$ snapshots and have used $\sigma_{\rm e}=0.3$ and 
$n_{\rm gal}=20$ galaxies/arcmin$^{2}$ to fix the error for each bin in the $\chi^{2}$-fitting metric.  
However we have not added shape noise (through Equation~\ref{eqn:staterr}) to the
reduced tangential shear profile in order to illustrate the intrinsic scatter 
and bias in the WL masses, as discussed in \S\ref{sec:fitmeth}. Also, for these fiducial results we
use 15 logarithmic bins in radius from the inner radial limit of 1
arcminute to an outer radial limit of $20$ arcminutes at $z=0.25$ and
10 logarithmic bins from 1 to 10 arcminutes at $z=0.50$.  At $z=0.25$ our choice of outer 
radial fit limit matches approximately that used by \citet{okabe2010} for 30 low-redshift clusters.  At $z=0.50$, we 
chose the outer radial fit limit of 10 arcminutes to match approximately that of the low-redshift halos relative to \rfivec. 

For our primary results, we fit for masses at an overdensity of
500 with respect to the critical density at the redshift of our halos.
Qualitatively, the intrinsic scatter and bias in the WL masses for other
overdensity mass definitions are similar.  The effects of halo shape
and orientation with respect to the LOS are discussed
in \S\ref{sec:haloshape}.  In \S\ref{sec:isbiased}, we give estimates
for the scatter and bias in WL mass estimates under changing
observational conditions, including varying overdensity fits, number
of bins, maximum fit radius, source galaxy densities, and source galaxy shape
noise.  We emphasize again that the results of this section are
specific to the method we have chosen for measuring the WL masses,
even when we do not refer explicitly to our method itself.
In \S\ref{sec:discuss} we will present a full discussion of our
results in comparison to previous work.

\subsection{Intrinsic Bias and Scatter in WL Mass Estimates}\label{sec:lssscat}

We will consider the distribution of WL masses at fixed 3D mass for a series of line-of-sight integration lengths in order to study the sources of bias and scatter.  We will use integration lengths of 6, 15, 30, 60, 120, 240, and 400 \hmpc.  An integration length of 400 \hmpc, for example, indicates that all the matter in the simulation from $-200$ \hmpc\ to $+200$ \hmpc\ along the LOS (with the cluster at zero) was included in constructing the reduced tangential shear profile.  For each simulation snapshot and integration length we obtain a WL mass estimate of \mfivec\ for the cluster under consideration and can compare this WL mass estimate to the \mfivec\ mass measured in three dimensions.   

Figure~\ref{fig:scatdist} shows the intrinsic scatter and bias in the
WL masses as a function of LOS integration length for the population
of halos at $z=0.25$ and $z=0.50$ snapshots. We have made two
different mass cuts at each redshift to illustrate the difference
between high- and low-mass halos.  For the $z=0.25$ snapshot, we have
kept the halos with \mfivec\ above $6.0\times10^{13}\hmsun$ and
$2.0\times10^{14}\hmsun$ for the low and high mass halo samples
respectively.  For the $z=0.50$ snapshot the cuts were placed at
$5.0\times10^{13}\hmsun$ and $1.5\times10^{14}\hmsun$.  These mass thresholds are set
so that the halo sample is complete above the low-mass threshold, so that there is 
approximately the same number of halos in the high-mass sample 
at both redshifts, and so that the qualitative differences in the shape of the scatter in the WL masses 
as a function of LOS integration length between the low- and high-mass halo samples are maximized.

The scatter, denoted as $\sigma_{\ln\mwl}$, is the width of the best fit Gaussian to the residuals from a fit of the $\ln\mwl$--$\ln\mfivec$ relation of the form
\begin{equation}\label{eqn:powlawform}
\ln\left(\frac{\mwl}{M_{p}}\right)=\beta+\alpha\ln\left(\frac{\mfivec}{M_{p}}\right)\ .
\end{equation}
The bias is defined as $\beta$ from the equation above and is the bias
in $\langle\ln\mwl|\mfivec\rangle$. $M_{p}$ is a pivot mass chosen so
that the errors on $\beta$ and $\alpha$ are uncorrelated.  $\alpha$ is
the slope of the relation.  The covariance matrix of the quantities
$\left\{\beta,\alpha,\sigma_{\ln\mwl}\right\}$ is computed using the
jackknife method described in \S\ref{sec:fitmeth}.  We have computed
all of the correlations of the quantities
$\left\{\beta,\alpha,\sigma_{\ln\mwl}\right\}$ between the various LOS
integration lengths and mass ranges using the jackknife method as
well. The error bars shown in Figure~\ref{fig:scatdist} are the
square-root of the diagonal elements of the jackknife covariance
matrix.  The best-fit parameters of Equation \ref{eqn:powlawform} for
each mass threshold and snapshot using the 400 \hmpc\ integration length
for \mfivec\ are given in Table \ref{table:fulllosscatter}.  The
scatter in the WL masses listed in this table is extrapolated to the
full LOS as described below.

In the method we adopt (i.e., the NFW profile fitting of the reduced tangential
shear profile), the WL masses are on average biased low by $\approx6\%$
in the $z=0.25$ and $z=0.50$ snapshots.    
A large portion, but not all, of the bias in the WL masses 
occurs because the NFW profile is a poor description of the actual shear 
profiles of clusters at the radii used in the fitting.
We illustrate this issue as follows.  For each halo we fit an NFW
profile to the halo's three-dimensional mass profile within \rfivec.  Then using the
parameters from this fit, we predict the reduced shear profile,
$g_{\mathrm{T,NFW}}$.  In Figure~\ref{fig:gbias} we have plotted the fractional difference between the true
reduced shear profile, $g_{\mathrm{T}}$, and the prediction from the
NFW fit of the three-dimensional halo mass profile,
$(g_{\mathrm{T}}-g_{\mathrm{T,NFW}})/g_{\mathrm{T,NFW}}$ for cluster halos in different ranges of mass \mfivec.  Note that
we compute the factional differences for each halo individually and
then average over the mass bin.  The lines at the top of the figure
indicate the value of \rfivec.  The virial radius of each halo is
$\approx2\rfivec$, which is well within 10 arcminutes for both analyzed redshifts for most clusters in the sample.  Typical fits of the reduced tangential shear
profile extend out to 10\arcmin-20\arcmin, where the actual reduced
shear $g_{\mathrm{T}}$ deviates significantly from the predictions of
the three-dimensional model. These deviations first become
more negative as the density profile becomes steeper than the NFW profile in the cluster infall region. 
At larger radii the NFW model asymptotes to zero and the fractional deviations then 
become more positive due to 2-halo contributions of matter around the cluster.

A simple Monte Carlo estimate of the bias in the WL masses from the
results of Figure~\ref{fig:gbias} shows that fitting an incorrect
model can account for $\approx2-4\%$ of the bias in the WL masses at
any overdensity and redshift.  We make this estimate as follows.  We
generate reduced tangential shear profiles using a given mass and
concentration of the cluster.  We then introduce a bias in these
profiles using the results of Figure~\ref{fig:gbias}.  Then we fit the
reduced tangential shear profile to determine a WL mass and compute
the fractional bias in the WL mass.  Our results indicate that using
an improved fitting function, or limiting the radial range of the fits
to the virial region, will decrease, but may not fully eliminate, the 
bias in our WL mass estimates.

There are other physical sources of bias in the WL masses, only some of 
which are included in Figure~\ref{fig:gbias}.  Substructure in the outskirts of halos 
will cause the shear profile to deviate low at radii less than the radius of the substructure and will 
cause the shear profile to deviate high at radii greater than the substructure.  This source of bias 
is already included in the estimates in Figure~\ref{fig:gbias} since we are comparing the shear with all 
substructures in the simulation included to the expected shear from the three-dimensional halo profile. 
Additionally, averaging the WL masses estimated from spherically-symmetric NFW model fits to individual triaxial halos with different orientations along the LOS can 
produce a biased average WL mass at the level of a few percent for a single set of axis ratios \citep{corless2007}. However 
when the WL masses from fits to individual clusters are averaged over the entire population of 
halos with a range of axis ratios the net bias is small, $\approx1\%$ high \citep{corless2007}.  In addition to the effects of halo shape, 
the correlation between the halo major axis and the direction to nearby halos and filaments can potentially change the expected 
bias in the WL masses from this orientation averaging effect as well.  Our simulations include all of these correlations automatically and so 
are a useful tool for studying these effects in aggregate.  

Note also that the bias in the WL masses depends sensitively on the fitting method.  
If we use the $\chi^{2}_{\rm uw}$ or $\chi^{2}_{\rm abs}$ fitting metrics, the WL masses are 
only biased low by $\approx3-4\%$.  These metrics tend to weigh the inner bins 
of the reduced tangential shear profile in the fit more than a pure $\chi^{2}$ fitting metric.
The reduced tangential shear profile is biased less in the inner regions in Figure~
\ref{fig:gbias}, so that the WL masses tend to be less biased using these alternative fitting metrics.

We additionally find that the exact choice of halo center can have a strong effect on 
the bias in the WL masses.  For example, if we allow the halo centers to move randomly $\pm100$ \hkpc\ comoving  
in the two orthogonal directions in the projected halo field about 
the fiducial center defined by the halo finder, the bias increases to $\approx7-10\%$ low for the 400 \hmpc\ integration length at $z=0.25$.  
We can also search for the peak of the convergence field for each halo.  We search the 
region defined by $\pm12$ grid cells about the grid center ($\pm469$ \hkpc\ in comoving distance) 
for the peak of the convergence field for the 400 \hmpc\ integration length at $z=0.25$.  
In this case, the peak tends to be approximately two grid cells (78 \hkpc\ comoving) or less away from the fiducial 
halo center defined by the halo finder.  The bias in the WL masses increases by $\lesssim1\%$ using these new halo centers.  Thus we can 
conclude that the halo centers defined by our halo finder are robust in this regard.  The effect of halo centering on the bias in the WL masses is quite sensitive to the exact choice of inner fitting radius as shown by \citet{hoekstra2010}.  We do not explore this dependence in this work.

\begin{deluxetable}{cccc}
\tablecolumns{4}
\tablecaption{Intrinsic Relation Between Weak Lensing Mass Estimates of \mfivec and 3D Mass.\tablenotemark{a}\label{table:fulllosscatter}}
\tablehead{\colhead{$M_{p}$} & \colhead{$\beta$} & \colhead{$\alpha$} & \colhead{$\sigma_{\ln\mwl}$\tablenotemark{b}}}
\startdata
\multicolumn{4}{c}{$z=0.25$, $\mfivec\geq6.0\times10^{13}\hmsun$} \\ \\
$9.74\times10^{13}\hmsun$ & $-0.059\pm0.003$ & $0.997\pm0.006$ & $0.297\pm0.008$ \\\\
\multicolumn{4}{c}{$z=0.25$, $\mfivec\geq2.0\times10^{14}\hmsun$} \\ \\
$2.80\times10^{14}\hmsun$ & $-0.052\pm0.007$ & $0.99\pm0.03$ & $0.207\pm0.005$ \\\\
\multicolumn{4}{c}{$z=0.50$, $\mfivec\geq5.0\times10^{13}\hmsun$} \\ \\
$7.89\times10^{13}\hmsun$ & $-0.063\pm0.003$ & $1.001\pm0.007$ & $0.311\pm0.009$ \\\\
\multicolumn{4}{c}{$z=0.50$, $\mfivec\geq1.5\times10^{14}\hmsun$} \\ \\
$2.07\times10^{14}\hmsun$ & $-0.054\pm0.008$ & $0.96\pm0.03$ & $0.22\pm0.01$
\enddata
\tablenotetext{a}{See Equation~\ref{eqn:powlawform} for definitions of the parameters $M_{p}$, $\beta$, and $\alpha$.}
\tablenotetext{b}{These numbers have been extrapolated using Equation~\ref{eqn:fracscatmodel}. See \S\ref{sec:lssscat} for details.}
\end{deluxetable}

The slope $\alpha$ of the relation in Equation~\ref{eqn:powlawform} is
generally consistent with unity at large LOS integration lengths.
However there are $\gtrsim3\sigma$ deviations of the slope below or
above unity in some mass ranges, especially at the smallest LOS
integration length in each snapshot for both mass bins.  A relation
with a slope significantly different than unity would indicate that
the bias in the WL masses depends on the halo mass.  However, as shown
in Figure~\ref{fig:gbias}, the deviation of the reduced tangential
shear profiles from the NFW model are nearly the same over all of the
mass bins, consistent with the slope of the \mwl-\mfivec\ being close
to unity.

As the LOS integration length is increased, so does the scatter in the
WL masses.  The scatter for the low-mass halo sample increases more
strongly than that for the high-mass halo sample in both snapshots.
Note that correlated structures at distances $\approx 3-10$ \hmpc\ contribute 
to the scatter. This correlated structure is not due to the triaxiality of the cluster 
itself, but is due to neighboring groups, clusters, and filaments. 
At distances $\gtrsim 10$ \hmpc, the scatter is generated from a 
superposition of many largely uncorrelated structures.  

We propose a simple toy model of the scatter as a function of mass and LOS integration
length which can explain the trends in Figure~\ref{fig:scatdist}.  We
assume that the scatter in the WL masses due to uncorrelated LSS is
proportional to the scatter in the shear.  We compute the fractional
scatter in the WL masses as
\begin{equation}\label{eqn:fracscatmodel}
\frac{\Delta\mwl}{\mwl} = \frac{\sqrt{\left[fM_{\rm med}\right]^{2} + \left[A\sigma_{\rm LSS}(d)\right]^{2}}}{M_{\rm med}}
\end{equation}
where $f$ is the fractional intrinsic scatter in the WL masses due to
halo triaxiality and correlated LSS, $\sigma_{\rm LSS}(d)$ is the
scatter in the shear due to uncorrelated LSS as a function of the LOS
integration length $d$, and $A$ is a proportionality constant with
units of \hmsun\ that is independent of the LOS integration length $d$
and halo mass.  $M_{med}$ is the median halo mass of the sample under
consideration.  The form of this model is motivated by the fact that
scatter from uncorrelated LSS should simply add in quadrature to the
scatter from triaxial halo shapes and correlated LSS.  We compute
$\sigma_{\rm LSS}(d)$ as a function of integration length using the
results of \citet{hoekstra2003} with the transfer function
from \citet{eisenstein1998} and the non-linear matter power spectrum
from \citet{smith2003}.  In general, $\sigma_{\rm LSS}(d)$ depends on
the radius and width of the annulus used to average the tangential
shear.  However, as shown in \citet{hoekstra2003}, this dependence is
not strong.  We use a bin of radius 10 arcminutes with a width much less
than the radius to get a representative value.  For each snapshot this
model is fit the to the last four LOS integration length points
(i.e. 60, 120, 240 and 400 \hmpc) for both mass bins simultaneously
using the jackknife covariance matrix computed above.  We assume that
both mass bins have the same value of $A$, but different values of
$f$, $f_{\rm low}$ and $f_{\rm high}$, for a total of three free parameters
per snapshot and five degrees of freedom in the fit.

\begin{figure*}[t]
\begin{center}
\includegraphics[scale=0.46]{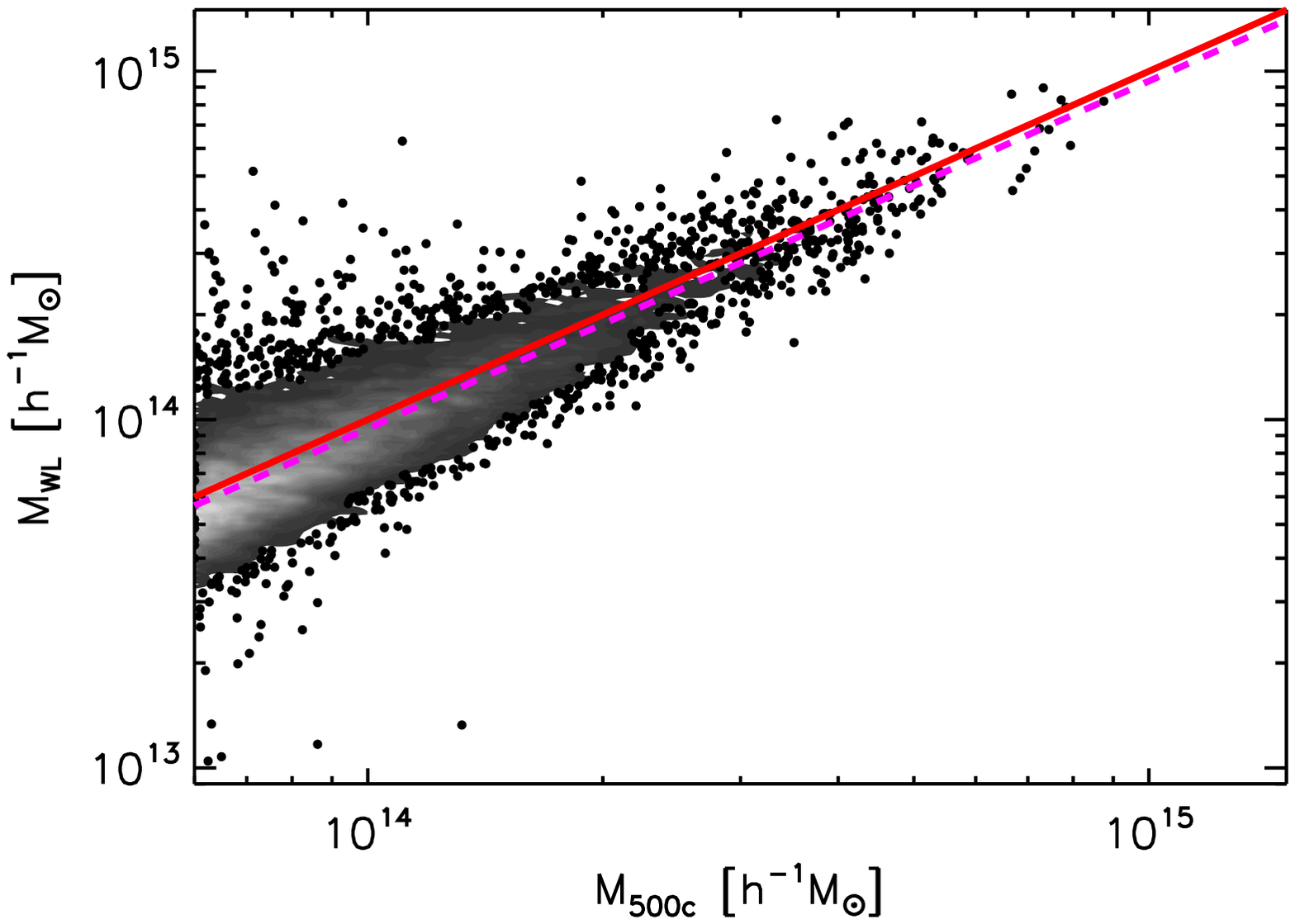}
\includegraphics[scale=0.46]{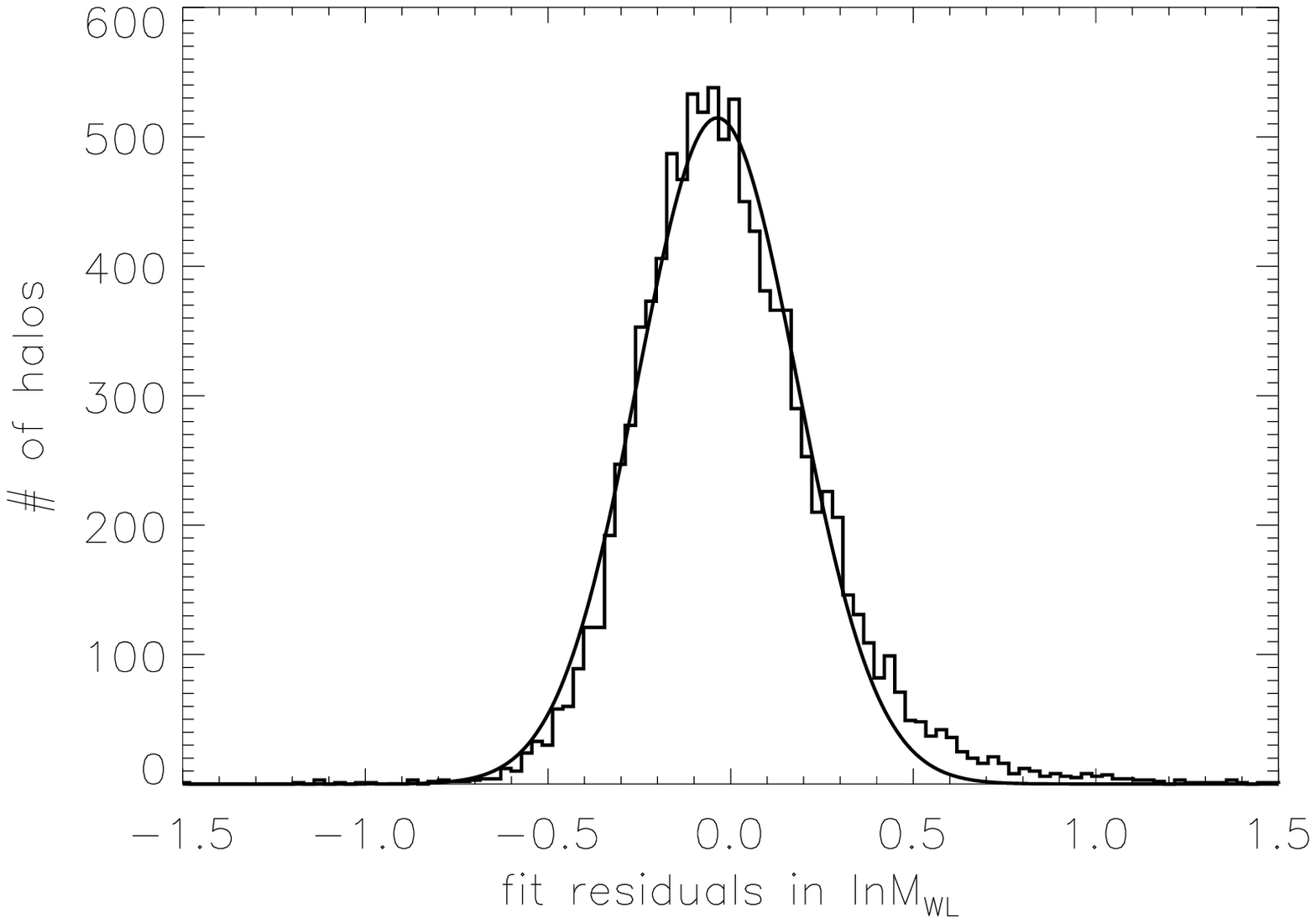}
\end{center}
\figcaption[]{\textit{Left}: Scatter-contour plot of \mfivec obtained from fitting the reduced tangential
  shear profile, \mwl, against the 3D \mfivec\ mass measured in
  the simulation.  The ``high-density'' regions of the plot are shown with
  grey shaded contours (light to dark indicates high- to low-density)
  and the rest of the data is shown as points.  The solid (red) line
  marks a one-to-one ratio and the dashed (magenta) line is the
  best-fitting power-law given in Equation~\ref{eqn:powlawform}.  The
  best-fit power law is biased low by $\approx6\%$ but the slope is
  consistent with unity. \textit{Right}: Residuals from the
  best-fitting power law on the left.  The solid line is the
  best-fitting Gaussian model to the residuals in $\ln\mfivec$.  The skewness of the
  residuals, $0.70\pm0.14$, is clearly non-zero indicating that the
  conditional distribution of \mwl\ at fixed \mfivec\ is close to, but
  not quite log-normal. See \S\ref{sec:lssscat} for
  details.\label{fig:ensmass}}
\end{figure*}

The fractional scatter predicted by Equation~\ref{eqn:fracscatmodel}
for the low- and high-mass sample of halos is shown as the solid lines
in Figure \ref{fig:scatdist}.  We have plotted model predictions for
points not used to fix the value of $A$, $f_{\rm low}$, and $f_{\rm
high}$ for the low mass bins in each snapshot as well.  While the
specific value of $A$ is somewhat arbitrary, the values of $f_{\rm
low}$ and $f_{\rm high}$ for each snapshot indicate the total amount
of scatter due to halo triaxiality and correlated LSS.  The three
parameters for the $z=0.25$ snapshot are
$A=(7.5\pm0.3)\times10^{15}\hmsun$, $f_{\rm low}=0.170\pm0.002$, and
$f_{\rm high}=0.189\pm0.006$.  For the $z=0.50$ snapshot the parameters
are $A=(6.6\pm0.3)\times10^{15}\hmsun$, $f_{\rm low}=0.163\pm0.003$, and
$f_{\rm high}=0.19\pm0.01$. Figure \ref{fig:scatdist} shows that about $80\%$ of 
the scatter $f$ is due to matter within $\approx 3$ \hmpc\ of clusters (i.e., within $2-3$ virial radii) 
and $\approx 20\%$ is due to the matter in correlated structures between 3 and 10 \hmpc. 

This model summarizes nicely the way in which uncorrelated LSS along
the LOS effects WL masses.  Uncorrelated LSS along the LOS adds
approximately the same amount of scatter to mass estimates independent of
halo mass.  Thus, in terms of fractional scatter, low-mass halos
are more strongly affected by uncorrelated LSS along the LOS because
they produce less shear as compared to high-mass halos.  This fact
motivates the assumption that parameter $A$ is the same for low- and
high-mass halos.

The differences between $f_{\rm low}$ and $f_{\rm high}$ for the
different cluster mass thresholds and snapshots are harder to explain
with a simple model.  They are essentially due to changes in halo bias
and shape as a function of mass.  Less massive halos tend to be more
spherical \citep[see e.g.,][]{allgood2006} than more massive halos.
Thus the fractional scatter generated by averaging over many
orientations is smaller for lower mass halos.  Additionally, higher
mass halos are more biased with respect to the dark matter than lower
mass halos (see \citeauthor{tinker2010} \citeyear{tinker2010}
for a recent study of halo bias).  Therefore, contributions to the
scatter in the WL masses from correlated structure outside the virial
radius (i.e. any slight increase in the scatter from an LOS
integration length of 6 to $\approx120$ \hmpc) along the LOS generally
are stronger for more massive halos.  Based on the changes in the
scatter as function of LOS distance in Figure~\ref{fig:scatdist}, the
effects of halo shape seem to be dominant in determining $f_{\rm low}$
and $f_{\rm high}$.  However, as we noted above, the effects of
correlated LSS outside the virial radius are not negligible.

Additionally, the model in Equation~\ref{eqn:fracscatmodel} can be used to extrapolate the 
scatter measured in the simulations over a restricted LOS to the full LOS.  We simply 
evaluate the model using the value of $\sigma_{\rm LSS}(d)$ from integrating over the 
entire LOS from the observer to the sources at redshift one.  The results are given in 
Table~\ref{table:fulllosscatter} for both snapshots and mass bins.  The extrapolated scatter 
values represent the total amount of intrinsic scatter in WL mass estimates of \mfivec\ for our 
fitting method from all sources along the LOS to redshift $z=1$.  Note that this extrapolation and the 
resulting effects of uncorrelated LSS on the scatter in the WL masses will change if the source galaxies 
extend to higher redshifts.

Finally, in Figure~\ref{fig:ensmass} we show a contour-scatter plot of
the $\ln\mwl$-$\ln\mfivec$ relation and a histogram of the residuals
about the best-fit power law for the $z=0.25$ snapshot at the
400 \hmpc\ integration length.  We have used all halos with
$\mfivec\geq6.0\times10^{13}\hmsun$ for this plot. There is a small
number of halos in the simulation where the WL mass differs from the
3D mass by a factor of five or more (the outliers in the left panel
of Figure~\ref{fig:ensmass}).  We have visually checked the reduced
shear profiles of these halos and found that they are negative in one
or more bins and/or are strongly non-monotonic.  These pathological
cases are due to the presence of a second mass peak in the shear field
which cancels shear from the main peak associated with the halo.

The conditional distribution of the WL masses from the best-fit mean 
relation shown in the right panel of Figure~\ref{fig:ensmass} is 
close to log-normal.  However there is a statistically significant tail at 
high WL masses.  The skewness of the distribution in the right panel 
of Figure~\ref{fig:ensmass} is $0.70\pm0.14$ for halos with $\mfivec\geq6.0\times10^{13}\hmsun$.  
For the higher mass sample with $\mfivec\geq2.0\times10^{14}\hmsun$, the conditional distribution 
is somewhat closer to log-normal with skewness $0.38\pm0.11$. The conditional distribution 
at $z=0.50$ is quite similar with skewness $0.43\pm0.18$ for the low mass sample 
with $\mfivec\geq5.0\times10^{13}\hmsun$ and $0.49\pm0.12$ for the high 
mass sample with $\mfivec\geq1.5\times10^{14}\hmsun$.

\subsection{The Effects of Halo Shape and Orientation}\label{sec:haloshape}

We have also investigated the correlation between the orientation of the halo with respect to the LOS and the deviation of its WL mass from the mean relation as a function of LOS integration length.  Specifically, we compute the correlation coefficient between $\cos\theta$ and $\Delta\ln\mwl=\ln\mwl-\left\langle\ln\mwl|\mfivec\right\rangle$.  Here $\theta$ is the angle of the halo's major axis with respect to the LOS (see \S\ref{sec:siminfo} for details) and $\left\langle\ln\mwl|\mfivec\right\rangle$ is the best-fit mean relation defined in Equation~\ref{eqn:powlawform}.  This correlation coefficient is $\approx0.68$ at the LOS integration length 6 \hmpc\ and declines to $\approx0.51$ at 400 \hmpc\ for the low mass halo sample defined above from the $z=0.25$ snapshot.  For the high mass halo sample from the same snapshot, the correlation coefficients at 6 and 400 \hmpc\ are $\approx0.72$ and $\approx0.61$ respectively.  The correlation coefficient between $\cos\theta$ and $\Delta\ln\mwl$ is smooth and monotonic from 6 to 400 \hmpc.  For the high mass halos, the correlation coefficient shows a slight increase from 3 to 6 \hmpc\ as the LOS integration length increases to include all of the halo's mass.  Similar trends are seen in the $z=0.50$ snapshot.  

The correlation coefficients decrease as the LOS integration length is
increased because of random perturbations to the shear profiles due to
uncorrelated LSS or due to imperfect alignment of correlated
structures.  The decline is stronger for lower mass halos because they
produce less shear, so that random perturbations have a larger effect.
In addition to the fact that smaller halos produce less shear, it is
known that the orientation of halos is correlated with the orientation
of the filamentary structure around them (see \S\ref{sec:intro} for
references).  Thus the filamentary, correlated LSS around larger
halos, which are more highly biased, will work to maintain the
correlation coefficient as the LOS integration length is increased.

\begin{deluxetable*}{ccccccccc}
\tabletypesize{\scriptsize}
\tablecolumns{9}
\tablecaption{Bias and Scatter in Weak Lensing Masses Estimates of \mfivec\ at $z=0.25$.\tablenotemark{a}\label{table:biases500c}}
\tablehead{\colhead{maximum fit radius} & \multicolumn{2}{c}{12 bins} & \multicolumn{2}{c}{15 bins} & \multicolumn{2}{c}{17 bins} & \multicolumn{2}{c}{20 bins}}
\startdata
&\multicolumn{8}{c}{$\{\sigma_{\rm e},n_{\rm gal}\}=\{0.3,10\}$} \\ \\
& Bias\tablenotemark{b} & Scatter\tablenotemark{c} & Bias & Scatter & Bias & Scatter & Bias & Scatter \\
15$\arcmin$ & -0.06$\pm$0.01 & 0.38$\pm$0.02& -0.06$\pm$0.01 & 0.39$\pm$0.02& -0.08$\pm$0.01 & 0.34$\pm$0.02& -0.08$\pm$0.01 & 0.35$\pm$0.02\\ 
20$\arcmin$ & -0.09$\pm$0.01 & 0.34$\pm$0.02& -0.10$\pm$0.02 & 0.37$\pm$0.04& -0.09$\pm$0.01 & 0.33$\pm$0.03& -0.10$\pm$0.01 & 0.36$\pm$0.03\\ 
25$\arcmin$ & -0.11$\pm$0.01 & 0.34$\pm$0.03& -0.10$\pm$0.01 & 0.37$\pm$0.02& -0.12$\pm$0.01 & 0.33$\pm$0.02& -0.10$\pm$0.01 & 0.35$\pm$0.03\\ \\ 
&\multicolumn{8}{c}{$\{\sigma_{\rm e},n_{\rm gal}\}=\{0.3,20\}$} \\ \\
& Bias & Scatter & Bias & Scatter & Bias & Scatter & Bias & Scatter \\
15$\arcmin$ & -0.06$\pm$0.01 & 0.29$\pm$0.01& -0.05$\pm$0.01 & 0.29$\pm$0.02& -0.06$\pm$0.01 & 0.28$\pm$0.02& -0.07$\pm$0.01 & 0.27$\pm$0.02\\ 
20$\arcmin$ & -0.07$\pm$0.01 & 0.27$\pm$0.02& -0.07$\pm$0.01 & 0.29$\pm$0.02& -0.08$\pm$0.01 & 0.29$\pm$0.02& -0.08$\pm$0.01 & 0.29$\pm$0.02\\ 
25$\arcmin$ & -0.09$\pm$0.01 & 0.29$\pm$0.02& -0.10$\pm$0.01 & 0.29$\pm$0.02& -0.09$\pm$0.01 & 0.27$\pm$0.02& -0.10$\pm$0.01 & 0.28$\pm$0.02\\ \\ 
&\multicolumn{8}{c}{$\{\sigma_{\rm e},n_{\rm gal}\}=\{0.3,40\}$} \\ \\
& Bias & Scatter & Bias & Scatter & Bias & Scatter & Bias & Scatter \\
15$\arcmin$ & -0.05$\pm$0.01 & 0.26$\pm$0.02& -0.05$\pm$0.01 & 0.26$\pm$0.02& -0.05$\pm$0.01 & 0.25$\pm$0.02& -0.05$\pm$0.01 & 0.27$\pm$0.01\\ 
20$\arcmin$ & -0.06$\pm$0.01 & 0.24$\pm$0.02& -0.07$\pm$0.01 & 0.25$\pm$0.02& -0.06$\pm$0.01 & 0.24$\pm$0.01& -0.07$\pm$0.01 & 0.23$\pm$0.02\\ 
25$\arcmin$ & -0.07$\pm$0.01 & 0.26$\pm$0.01& -0.07$\pm$0.01 & 0.22$\pm$0.02& -0.08$\pm$0.01 & 0.24$\pm$0.01& -0.08$\pm$0.01 & 0.25$\pm$0.01
\enddata
\tablenotetext{a}{We have used all halos with $\mfivec\geq2.0\times10^{14}\hmsun$. Additionally, we require that the WL mass differ from the 3D mass by no more than a factor of five.}
\tablenotetext{b}{This quantity is $\beta$ from Equation~\ref{eqn:powlawform} and is the bias in $\langle\ln\mwl|M_{\Delta}\rangle$.}
\tablenotetext{c}{The scatter is defined as width of the best-fit Gaussian to the residuals of the \mwl-$M_{\Delta}$ relation, $\sigma_{\ln\mwl}$.}
\end{deluxetable*}

It is interesting that even considering all matter between $-200$ and
$+200$ \hmpc, the correlation coefficients are still substantial,
$\approx0.50-0.60$.  We expect the correlation
coefficients to decrease more as the LOS integration length is
increased, but extrapolating accurately to the final values for a full
LOS from the simulation data we have presented here is difficult.
However, we can do this extrapolation approximately as follows.
We use the toy model proposed in Equation~\ref{eqn:fracscatmodel} above 
to include the effects of uncorrelated LSS along the LOS on the WL masses.
The term proportional to $(A\sigma_{\rm LSS}(d)/M_{\rm med})^{2}$ in Equation~\ref{eqn:fracscatmodel} 
is the extra variance introduced into the WL masses due to the effects of 
uncorrelated LSS projections along the LOS.  We thus use
log-normal random deviates with variance equal to $(A\sigma_{\rm
LSS}(d)/\mfivec)^{2}$ and zero mean to add scatter to the WL masses in order to simulate the effects of LSS projections 
along the LOS not included in our simulation box.  Here we use the value of $\sigma_{\rm LSS}(d)$ 
which corresponds to the extra scatter due to a full LOS from lensing sources to the observer and \mfivec\ is set to the 3D mass of 
each cluster individually when adding the random scatter. 
We start with the WL masses at an integration length of 120 \hmpc.    
The choice of starting integration length is motivated by the fact that most
of the correlated structure along the LOS due to filamentary LSS is
within 60-100 \hmpc\ of the halo center.  We then recompute 
the correlation coefficients.  We find a correlation coefficient of
$\approx0.37$ for the low-mass bins in both the $z=0.25$ and $z=0.50$
snapshots.  For the high-mass bins in both snapshots, the correlation
coefficients are $\approx0.58$.  If we use the WL masses at an
integration length of 240 \hmpc, these numbers change by
$\lesssim0.01$, indicating that our extrapolation is robust to this
choice. 

While there is certainly some uncertainty in this extrapolation, it is
clear that the correlation still remains positive even integrating
over a full LOS.  Stated differently, the key point is that the {\it
sign} (i.e. positive or negative) of the deviation of the WL mass from
the 3D mass is on average set by the orientation of the matter
within and near the virial radius of the halo.  Matter outside the
virial radius of the halo along the LOS changes the strength of the
correlation between the orientation of the halo and the deviations of
the WL masses from the 3D masses, but the correlation is always
positive at high significance on the scales probed by our simulation
and remains positive even after extrapolating to a full LOS.

In addition to halo orientation with respect to the LOS, halo shapes
also influence the WL masses. To investigate this effect, we build
subsamples of halos which are more spherical or more triaxial based on
the minor to major axis ratio, denoted here as $S$
(see \S\ref{sec:siminfo} for details concerning its computation). We
fit a power-law to the mean value of $S$ as a function of \mfivec\ (a
similar relation was used
in \citeauthor{allgood2006} \citeyear{allgood2006}). Using the
mean power-law relation between $S$ and \mfivec, we compare all halos
with $S$ in the upper quartile of the distribution of $S$ at fixed mass 
(i.e. the more spherical halos) to the entire halo population and all
halos with $S$ in the lower quartile of the distribution of $S$ at fixed mass (i.e. the more triaxial halos) 
to the entire halo population using the same mass bins as before.  These cuts generate four halo samples per
snapshot.

We find that while the bias in normalization and slope of
the \mwl-\mfivec\ relations are unaffected in a statistically
significant way by cuts on halo shape, the scatter in the WL masses does
depend on halo shape.  Not unexpectedly, the WL masses of more
spherical halos have less scatter than those of more triaxial halos.
In both snapshots for the low-mass cuts, the more triaxial halos have
$\approx3-5\%$ more scatter than the entire
halo population while the more spherical halos have $\approx3-5\%$
less scatter than the entire halo population.  For the high-mass cuts the 
mean shifts in the scatter between the different samples are the same, 
but the trend is not statistically significant given the jackknife errors. 
We find additionally that the difference in the scatter
in the WL masses between the more spherical halos and the entire
population increases marginally with integration length from 60 \hmpc\
to 400 \hmpc\ in the $z=0.50$ snapshot.  
Given that correlations between
halo orientation and LSS persist out to $\sim100$ \hmpc\ and that
spherical halos tend to be more highly biased (i.e. more clustered)
than triaxial halos at fixed mass \citep{faltenbacher2010} and thus
tend to dominate their local environment more strongly, this trend is
physically plausible. 

\subsection{Scatter and Bias Under Varying Observational Conditions}\label{sec:isbiased}

In this section we present the bias and scatter in the WL masses at
fixed 3D mass including the effects of galaxy shape noise.  For each halo
in the simulation, we vary the source density, amount of shape noise, maximum
radius of the fit, and the number of bins used in the fit.  We
consider WL mass estimates of \mtwom, \mtwoc, and \mfivec.  For each
set of observational parameters and mass overdensity definition, we measure the bias and scatter in the WL masses.  Since we are now including
observational errors in the reduced tangential shear profiles, the shear of some
of the lower mass halos will be undetectable.  We thus focus only on the highest mass halos in the
simulation which have the highest signal-to-noise.  

Additionally, we find that in the presence of observational errors, even while focusing on the high-mass halos only, there are still outliers in the WL masses, especially for the poorest observations with $\{\sigma_{\rm e},n_{\rm gal}\}=\{0.3,10\}$. These outliers generally occur when the signal-to-noise is low so that the WL mass differs from the 3D mass by a factor of five or more and is biased low.  We thus cut all halos where the WL mass differs from the 3D mass by a factor of five or more before computing the bias and scatter in the WL masses.  These cuts reject low signal-to-noise observations while still retaining a sample with a sharp mass threshold above which the sample is nearly $100\%$ complete.  We do not use direct cuts on signal-to-noise since these cuts result in a sample with varying completeness as a function of mass.  This extra cut has a negligible effect on the results at $z=0.25$, but does reduce the measured bias in the WL masses at $z=0.50$.

The bias and scatter in the WL mass measurements of \mfivec\ averaged
over all halos with $\mfivec\geq2.0\times10^{14}\hmsun$ at $z=0.25$ are given in Table
\ref{table:biases500c}.  We show results for different outer radial limits of
the fit and the number
of bins for the errors typical for a wide-area DES-like, deep ground-based, and
space-based observational surveys.  The scatter in the WL mass varies as a function of
the quality of the observations, but varies very little with the exact choice of outer radial fit limit or number of bins.  
The expected scatter in WL mass estimates of
\mfivec\ at fixed 3D mass for very common ground-based or DES-like observations is $\approx33-39\%$.
  For current common ground-based or DES-like observations, 
the dominant source of scatter is shape noise from background galaxies.   For deeper ground-based observations, 
the scatter drops slightly to $\approx27-29\%$.  For high-quality observations like those expected from LSST or 
space-based instruments, the scatter drops to $\approx22-27\%$.
As the number density of sources approaches that expected 
from space-based observations or LSST, the contribution to the scatter in the WL masses from galaxy shape noise
becomes comparable or even sub-dominant to the intrinsic scatter in the WL masses at $z=0.25$ for the radial fitting ranges we have considered here. If the radial fitting range is decreased significantly (e.g. to 10 arcminutes), the scatter in the WL masses will increase even for the highest quality observations simply because many fewer galaxies are used in the measurement and thus the signal-to-noise is lower.

\begin{deluxetable}{ccc}
\tabletypesize{\scriptsize}
\tablewidth{0.8\columnwidth}
\tablecolumns{3}
\tablecaption{Bias and Scatter in Weak Lensing Masses Estimates of \mtwom, \mtwoc, and \mfivec at $z=0.25$.\tablenotemark{a,b}\label{table:biasesalldef141}}
\tablehead{\colhead{mass definition $\Delta\rho(z)$\tablenotemark{c}} & \colhead{Bias\tablenotemark{d}} & \colhead{Scatter\tablenotemark{e}}}
\startdata
&\multicolumn{2}{c}{$\{\sigma_{\rm e},n_{\rm gal}\}=\{0.3,10\}$} \\ \\
$200\rho_{m}(z)$ & \phs 0.00$\pm$0.02 & 0.40$\pm$0.02\\ 
$200\rho_{c}(z)$ & -0.06$\pm$0.01 & 0.35$\pm$0.01\\ 
$500\rho_{c}(z)$ & -0.10$\pm$0.02 & 0.37$\pm$0.04\\ \\ 
&\multicolumn{2}{c}{$\{\sigma_{\rm e},n_{\rm gal}\}=\{0.3,20\}$} \\ \\
$200\rho_{m}(z)$ & \phs 0.02$\pm$0.01 & 0.32$\pm$0.02\\ 
$200\rho_{c}(z)$ & -0.06$\pm$0.01 & 0.27$\pm$0.02\\ 
$500\rho_{c}(z)$ & -0.07$\pm$0.01 & 0.29$\pm$0.02\\ \\ 
&\multicolumn{2}{c}{$\{\sigma_{\rm e},n_{\rm gal}\}=\{0.3,40\}$} \\ \\
$200\rho_{m}(z)$ & \phs 0.02$\pm$0.01 & 0.27$\pm$0.02\\ 
$200\rho_{c}(z)$ & -0.04$\pm$0.01 & 0.25$\pm$0.01\\ 
$500\rho_{c}(z)$ & -0.07$\pm$0.01 & 0.25$\pm$0.02
\enddata
\tablenotetext{a}{We have used all halos with $\mtwom\geq4.0\times10^{14}\hmsun$, $\mtwoc\geq3.0\times10^{14}$, and $\mfivec\geq2.0\times10^{14}$. Additionally, we require that the WL mass differ from the 3D mass by no more than a factor of five.}
\tablenotetext{b}{The WL masses were fit with an outer radial fit limit of 20 arcminutes using 15 bins.}
\tablenotetext{c}{$\Delta\rho(z)$ is defined through the relation $\mathrm{M}_{\Delta}=\Delta\rho(z)\frac{4}{3}\pi r^{3}$.}
\tablenotetext{d}{This quantity is $\beta$ from Equation~\ref{eqn:powlawform} and is the bias in $\langle\ln\mwl|M_{\Delta}\rangle$.}
\tablenotetext{e}{The scatter is defined as width of the best-fit Gaussian to the residuals of the \mwl-$M_{\Delta}$ relation, $\sigma_{\ln\mwl}$.}
\end{deluxetable}

\begin{deluxetable}{ccc}
\tabletypesize{\scriptsize}
\tablewidth{0.8\columnwidth}
\tablecolumns{3}
\tablecaption{Bias and Scatter in Weak Lensing Masses Estimates of \mtwom, \mtwoc, and \mfivec at $z=0.50$.\tablenotemark{a,b}\label{table:biasesalldef124}}
\tablehead{\colhead{mass definition $\Delta\rho(z)$\tablenotemark{c}} & \colhead{Bias\tablenotemark{d}} & \colhead{Scatter\tablenotemark{e}}}
\startdata
&\multicolumn{2}{c}{$\{\sigma_{\rm e},n_{\rm gal}\}=\{0.3,10\}$} \\ \\
$200\rho_{m}(z)$ & \phs 0.03$\pm$0.03 & 0.52$\pm$0.07\\ 
$200\rho_{c}(z)$ & -0.04$\pm$0.02 & 0.57$\pm$0.03\\ 
$500\rho_{c}(z)$ & -0.11$\pm$0.02 & 0.51$\pm$0.04\\ \\ 
&\multicolumn{2}{c}{$\{\sigma_{\rm e},n_{\rm gal}\}=\{0.3,20\}$} \\ \\
$200\rho_{m}(z)$ & -0.04$\pm$0.03 & 0.44$\pm$0.06\\ 
$200\rho_{c}(z)$ & -0.08$\pm$0.02 & 0.42$\pm$0.05\\ 
$500\rho_{c}(z)$ & -0.10$\pm$0.01 & 0.40$\pm$0.03\\ \\ 
&\multicolumn{2}{c}{$\{\sigma_{\rm e},n_{\rm gal}\}=\{0.3,40\}$} \\ \\
$200\rho_{m}(z)$ & \phs 0.02$\pm$0.02 & 0.32$\pm$0.08\\ 
$200\rho_{c}(z)$ & -0.06$\pm$0.01 & 0.36$\pm$0.03\\ 
$500\rho_{c}(z)$ & -0.09$\pm$0.01 & 0.33$\pm$0.02
\enddata
\tablenotetext{a}{We have used all halos with $\mtwom\geq4.0\times10^{14}\hmsun$, $\mtwoc\geq2.5\times10^{14}$, and $\mfivec\geq1.5\times10^{14}$. Additionally, we require that the WL mass differ from the 3D mass by no more than a factor of five.}
\tablenotetext{b}{The WL masses were fit with an outer radial fit limit of 10 arcminutes using 10 bins.}
\tablenotetext{c}{$\Delta\rho(z)$ is defined through the relation $\mathrm{M}_{\Delta}=\Delta\rho(z)\frac{4}{3}\pi r^{3}$.}
\tablenotetext{d}{This quantity is $\beta$ from Equation~\ref{eqn:powlawform} and is the bias in $\langle\ln\mwl|M_{\Delta}\rangle$.}
\tablenotetext{e}{The scatter is defined as width of the best-fit Gaussian to the residuals of the \mwl-$M_{\Delta}$ relation, $\sigma_{\ln\mwl}$.}
\end{deluxetable}

The bias in the WL masses at fixed 3D mass is in the range of
$[-12\%,-5\%]$ in all cases at $z=0.25$.  The errors in the bias from the jackknife samples are typically $\approx\pm1\%$
for the high-mass halo sample we are considering.  
The bias generally increases with increasing outer radial fit limit. 
The main cause of this bias is apparent in Figure~\ref{fig:gbias}: the deviation of the halo's true tangential shear profile from the 
NFW model we are using in our fitting method increases as the outer fit limit is increased, resulting in more bias in the WL masses.  We have confirmed this trend using the Monte Carlo method described above. 

For reference, the bias and scatter in WL mass measurements for various other mass
definitions at $z=0.25$ are given in Table \ref{table:biasesalldef141}.  For this
table we have set the outer radial limit of the fit to 20 arcminutes
and the number of bins to 15.  The shape noise contribution to the scatter in the WL masses is dominant for all mass definitions.  
The scatter increases as the overdensity decreases because smaller overdensities pivot the fit of the tangential shear profile away from the median observed radius \citep{okabe2010}.  The shift in the pivot point causes uncertainty in the measured concentration of the NFW profile to project into the WL mass, causing the scatter to increase.

The bias and scatter in the WL masses using an outer fit limit of 10 arcminutes 
and 10 bins for $z=0.50$ are given in Table~\ref{table:biasesalldef124}.  
The biases in the WL masses in the presence of shape noise for common ground-based 
observations are marginally larger at $z=0.50$ than at $z=0.25$.  Also there is 
significantly more scatter in the WL masses at $z=0.50$, due to the decreased radial 
range of the fits and the corresponding drop in signal-to-noise of the WL mass measurements.

Weak lensing masses measured at an overdensity of $\Delta\rho(z)=200\rho_{m}(z)$
seem to be unbiased at both $z=0.25$ and $z=0.50$.  Additionally, if we use 10 bins and an outer fit limit of 10
arcminutes for the $z=0.25$ halos, then the WL masses measured at
$\Delta\rho(z)=500\rho_{c}(z)$ are nearly unbiased, $-0.01\pm0.01$ for 
$\{\sigma_{\rm e},n_{\rm gal}\}=\{0.3,40\}$, as one would expect
from Figure~\ref{fig:gbias}.  However, with the 10 arcminute outer fit
limit for the $z=0.25$ halos, the WL masses measured at lower overdensities are then biased
high, $0.06\pm0.01$ for \mtwoc\ and $0.10\pm0.02$ for \mtwom\ with 
$\{\sigma_{\rm e},n_{\rm gal}\}=\{0.3,40\}$. The apparent discrepancies between 
these results and Figure~\ref{fig:gbias} occur because the NFW fits of the three-dimensional 
mass profile were done within \rfivec\ only.  Thus while Figure \ref{fig:gbias} 
reflects the differences between the observed shear profiles and the NFW prediction for \mfivec, 
NFW fits of the three-dimensional mass profile within \rtwoc\ or \rtwom\ would be different, and thus this figure for the lower
over density mass definitions would change.  Thus Figure~\ref{fig:gbias} 
is only applicable to the biases in WL estimates of \mfivec.

\subsection{Tests of a Iterative Strategy for Mitigating Biases in WL Mass Estimates}

As discussed in the previous section, the biases in the WL masses
depend principally on how far in radius the fit to the reduced tangential
shear profile is carried out. Given that the source of the bias are deviations of the actual mass
distribution from the NFW form outside the virial radius of the halo, one may wonder then whether the bias can
be removed by restricting the fits to radii within the virial radius only. We
have tested a simple algorithm of iteratively determining the outer
radial fit limit from the measured WL mass.  Specifically, we first
fit the reduced tangential shear profile within some fiducial outer
fit radius, 20 arcminutes for halos at $z=0.25$ and 10 arcminutes for
halos at $z=0.50$.  We then use the WL mass estimate resulting from such fit to refit
the reduced tangential shear profile within the virial radius
from the previous fit (i.e., \rfivec\ for \mfivec, \rtwom\ for
\mtwom, etc.).  This process is repeated until the WL mass converges, usually after a few iterations.  

For halos at $z=0.25$ above the mass limits listed in Table
\ref{table:biasesalldef141}, the bias in the WL masses over all halo
definitions and observational conditions using such an iterative fitting
method is in the range of $[-10\%,+8\%]$ with a $1\sigma$ error in
the range of $2-3\%$.  The average absolute bias in units of the
jackknife errors is $2.6\sigma$ for the iterative fitting method as
compared to $4.2\sigma$ for the biases listed in Table
\ref{table:biasesalldef141} (a fixed outer radial fit limit of 20
arcminutes).  Note that this improvement is driven mostly by the
increase in the jackknife errors and corresponding increases in the
scatter in the WL masses, since the average absolute biases for these
two methods are nearly identical around $\sim 5\%$.  

The iterative method does improve the estimate of masses defined within radii enclosing relatively high overdensity. Thus, for \mfivec\ at
$z=0.25$ and $n_{\rm gal}\gtrsim20$, the iterative fitting method results in  average bias of only $-2.5\pm 2.0\%$, i.e. no significant bias.  This is consistent with the results in the previous section
for WL mass estimates of \mfivec\ with an outer radial fit limit of 10 arcminutes at $z=0.25$.  The results for lower source densities with
\mfivec\ at $z=0.25$ are somewhat inconclusive because error bars on the mean bias are considerably large in this case. Due to 
large errors, the bias in \mfivec, $-10\pm 3\%$ is still consistent with zero for such source densities.

At $z=0.50$ for halos above the mass limits listed in Table
\ref{table:biasesalldef124}, the iterative method shows only a small improvement 
 over results for a fixed outer fit limit of 10 arcminutes, even for
\mfivec.  We find that \mfivec\ is biased by the same amount on average over all 
source densities, $-10\pm2\%$, for both the iterative method and for the fixed outer fit limit.
Given that Figure~\ref{fig:gbias} shows that model NFW tangential shear profiles
should be unbiased at $r\lesssim r_{500c}$, the bias must be due to
convergence of the method to incorrect $r_{500c}$ radius.  Indeed,
given that we start the method with profile fitted within
10 arcminutes, at which the model profiles are highly biased,
convergence to the correct radius is not guaranteed. If a measurement of \mfivec\ 
is the goal, the WL mass measurements should therefore be
restricted to even smaller radii to avoid biases, even though this will
result in degradation of the signal-to-noise of the measurements.  
For \mfivec, we find that by starting the iterative fitting method at 5 arcminutes, 
the biases drop to $-2\pm2\%$ on average over all source densities.
For masses defined within radii enclosing over densities lower than $500c$,
the iterative fitting results in biased mass estimates with the bias
in the range of $[-13\%,+7\%]$ with a $1\sigma$ error of $2-5\%$. 
While the bias is consistent with zero according to the error bars, the errors bars are large enough 
that we cannot distinguish changes in the biases at high precision. 
These results are thus inconclusive for over densities lower than $500c$.

Overall our results indicate that the bias in the WL masses can be alleviated for \mfivec\
measurements by restricting fits to smaller angular scales, and
iteratively decreasing the outer fit radius to converge to $r_{500c}$,
although care must be taken to perform fits only within radii where
NFW model is expected to be an unbiased model for the shear ($\lesssim
10^{\prime}$ at $z=0.25$ and $\lesssim 5^{\prime}$ at $z=0.5$, see
Fig.~\ref{fig:gbias}). However, this method does not eliminate biases
for other mass definitions enclosing lower overdensities and other
methods will need to be explored to eliminate bias in mass
measurements within larger radii.

\section{Discussion}\label{sec:discuss}

Our results presented in the previous section show that contributions
to the scatter and bias in WL masses estimated from an NFW fit comes 
from three physically distinct sources: matter within
the halo virial radius, correlated LSS at distances 6-20 \hmpc\ from
clusters, and uncorrelated LSS at larger distances.  Previous studies
have used a combination of analytic models and simulations to study
these different sources separately, while we have considered the
effects of all three simultaneously. In the subsections below, we
summarize the contributions of each of these sources and compare to 
various previous studies of this subject.  We also discuss in detail some implications 
of our results for studies of cosmology with galaxy clusters.

\subsection{Matter Within the Halo Virial Radius}

The matter within the virial radius and immediately outside of it is a
significant source of scatter and the main source of biases in the WL
mass estimates using NFW fits.  Specifically, the bias shown in
Figure~\ref{fig:scatdist} changes negligibly once the LOS integration
length is increased beyond 6 \hmpc.  The main origin of the bias is shown in
Figure~\ref{fig:gbias}, which shows that deviations of the mean
reduced tangential shear profile from NFW profile are significant
outside the virial radius. So when these radii are included in the NFW
fit, the resulting mass is biased low. The deviations in
Figure~\ref{fig:gbias} are consistent with the results of
\citet{tavio2008}, who have systematically studied density profiles of
halos beyond the virial radius.  

In addition to demonstrating that using an inaccurate density profile
can bias WL mass estimates, we have demonstrated that even if the
correct halo profile is known, there are still biases in the WL masses
which depend on the specific details of the fitting method and need to
be calibrated in simulations. Note that high-resolution simulations naturally 
include other sources of bias like substructure, halo triaxiality, and potential 
halo centering issues as well. Specifically, we have demonstrated in our simulations that
halo centering errors can introduce $\approx5\%$ negative biases in WL masses 
as has been seen before using analytic models \citep[see e.g.,][]{hoekstra2010}.  
The myriad of complications involved in WL mass estimation makes detailed studies
of shear fitting using shear fields derived from cosmological simulations indispensable in estimating
the bias and scatter of the weak lensing mass measurements.

The orientation of the major axis of halo mass distribution also affects the 
magnitude and sign of the bias in the WL mass estimates.  This effect
was discussed by \citet{clowe2004} for a small sample of simulated clusters using all
matter within 7.5 \hmpc\ of the halo center.  \citet{meneghetti2010a}
have detected this effect with a similarly small sample of halos using all
matter within 10 \hmpc\ of the halo center. Finally, \citet{oguri2005} and \citet{corless2007} used
analytic triaxial NFW models to arrive at a similar conclusion for
matter within the halo virial radius.  We have extended this type of
analysis to the full LOS, showing that these correlations persist to large distances, and 
to a much larger sample of simulated halos.

Halos in the roundest quartile of the distribution of $S$ at fixed mass 
have less scatter in the WL masses by $\approx3-5\%$ and halos in the 
most triaxial quartile of the distribution of $S$ at fixed mass
have $\approx3-5\%$ more scatter than the overall halo population. 
A similar conclusion for matter just associated with the halo itself was
found by \citet{corless2007} using analytic triaxial NFW models of halos.
\citet{marian2010} found using a different WL mass estimator that WL masses for low-mass halos have
less scatter than for high-mass halos.  They interpreted this effect
as due to the decrease of triaxiality with decreasing halo mass
expected in $\Lambda$CDM cosmology \citep[e.g.,][]{allgood2006}.  We
have presented a similar interpretation of our results in
Figure~\ref{fig:scatdist}. Finally, as indicated in
Figure~\ref{fig:scatdist}, the majority of the scatter in the WL
masses due to matter correlated with the halo is set by matter within
approximately two virial radii. \citet{marian2010} reached a similar conclusion 
for a different WL mass estimator.

\subsection{Correlated LSS}

For our WL mass measurement method, correlated LSS at distances
$\approx3-20$ \hmpc\ has a small, but non-negligible contribution
($\approx20\%$ of the total) to the scatter of WL masses (and no
effect on the bias). \citet{clowe2004} used the same WL mass estimator
as this study and saw hints of similar effects of triaxiality and
correlated LSS on WL masses to those we identify in this study.
However, given the small number of clusters analyzed, they
could not quantify the effect of correlated LSS on the scatter in the
WL masses.  \citet{metzler2001} found larger effects on both the bias
and the scatter in WL mass estimates due to correlated LSS. Similarly,
\citet{marian2010} found somewhat smaller effect on the scatter in the
WL masses due to correlated LSS than the ones we find in this study,
though they use a friends-of-friends halo finder which complicates the
separation of the effects of correlated LSS and triaxial halo shapes.
However, these differences are most likely due to differences in the
method used to estimate the WL masses: \citet{metzler2001} use an
aperture mass estimator, whereas \citet{marian2010} use a compensated
aperture mass estimator.  The comparison of these works with our own
clearly demonstrates that the properties of WL masses strongly depend
on how they are estimated.

Note also that in some of the previous studies the effects of correlated
LSS on WL masses have been studied in less direct ways. For example,
\citet{king2001a} studied a particular configuration of two halos in
close projection with a varying impact parameter.  They found changes
in the recovered WL masses that are similar to the scatter in the WL
masses measured from our simulations.  Halos identified using the
spherical overdensity algorithm used in our study are known to have
nearby, overlapping neighbors even at high masses
\citep[e.g.,][]{evrard2008}. This effect is the flip side of the well-known
``bridging effect'' in friends-of-friends halo finders, which often join such
neighboring halos into a single structure. The study of
\citet{king2001a} thus provides some insight into the effects of different
configurations of neighboring halos with respect to the LOS and origin
of scatter due to nearby, correlated large-scale structures.

\citet{deputter2005} have used a smaller 300 \hmpc\ simulation tiled 
along the LOS to study the total amount of scatter introduced in WL
masses due to LOS projections.  They estimate the scatter in WL masses
by computing the scatter in the tangential shear at a fixed radius due
to LSS projections.  While these authors do not distinguish between
correlated and uncorrelated LSS, given the high masses of the halos
they consider, we have demonstrated that the effects they observe are
due mostly to halo shape and correlated LSS.

\subsection{Uncorrelated LSS}

As the LOS integration length is increased into the regime of
uncorrelated LSS, we have found that for our WL mass estimator the
scatter increases due to random projections.  Additionally, we have
demonstrated that the model of \citet{hoekstra2003,hoekstra2001} based
on cosmic shear computations can correctly predict the increase of the
scatter with LOS integration length in this regime.  \citet{hoekstra2010a} have reached a similar 
conclusion using a fixed analytical NFW cluster model superimposed on top of uncorrelated LSS noise 
from ray-tracing through a large $N$-body simulation.  The formalism of \citet{hoekstra2003,hoekstra2001} 
also correctly predicts the different behavior of WL masses measured for low- and high-mass halos in the presence of random
projections along the LOS.  The scatter in the WL masses of low-mass
halos increases more than for high-mass halos as a function of LOS
integration length because the high-mass halos generate more shear
than the low-mass halos.  \citet{hoekstra2010a} find that uncorrelated LSS has a larger 
effect on the scatter in the WL masses than we find here.  They have sources out to $z=3$, so that they 
integrate over more mass fluctuations along the LOS than we do with sources at $z=1$.
This change in source redshift accounts approximately for these 
differences.  These differences also indicate 
that the relative contributions to the intrinsic scatter 
of triaxial halo shapes and correlated LSS versus uncorrelated LSS will change as the source redshift is increased, 
with the effects of uncorrelated LSS becoming stronger.
\citet{marian2010} find that uncorrelated LSS projections have a negligible effect on their WL masses estimated
with a compensated aperture mass filter.  Comparing their work with
our results, it is clear that their WL mass estimator is more
efficient than the one considered here in filtering out the effects of
uncorrelated LSS projections.  These differences highlight the need to
study each WL mass estimator individually in simulations in order to
understand its properties.

In addition to the intrinsic effects of matter along the LOS, other
authors have considered the effects WL shape noise as well
\citep{hoekstra2001,hoekstra2003,king2001a,king2001b,corless2007,meneghetti2010a,hoekstra2010a}.
While our estimate of the effects of shape noise are consistent with
results of these studies, we are also able to accurately compare the
effects of matter projections along the LOS with shape noise.  In
particular, we find that shape noise is a dominant source of scatter
in WL masses for most common ground-based observations
(i.e. $\{\sigma_{\rm e},n_{\rm gal}\}=\{0.3,10\}$ or $\{\sigma_{\rm e},n_{\rm
gal}\}=\{0.3,20\}$), even in the presence of triaxial halo shapes and
uncorrelated projections of mass along the LOS.  \citet{oguri2010} found that shape noise 
is the dominant source of scatter in WL masses estimated with a different method using 
ground-based observations of individual clusters as well.  As the weak lensing observations
become better, for the WL mass estimator considered here, shape noise
effects and intrinsic scatter will make comparable contributions to
the total scatter in the WL masses (see \citeauthor{clowe2004} \citeyear{clowe2004} and \citeauthor{hoekstra2010a} \citeyear{hoekstra2010a} for similar conclusions). 
In order to achieve precision calibration of cluster mass-observable relations for precision
cosmology any source of scatter and bias at the level of $\gtrsim
1-10\%$ needs to be considered and controlled.

\subsection{Implications for Scatter in Weak Lensing Mass-Observable Relations}\label{sec:wlm}

Recently, samples of galaxy clusters with masses inferred from weak lensing 
have grown to the point where scaling
relations between cluster observables and WL masses 
can be analyzed in detail \citep[e.g.,][]{zhang2008,vikhlinin2009b,henry2009,okabe2010b}.  For example,
\citet{okabe2010b} constrain both the normalization of and scatter in scaling relations between X-ray observables and WL masses
using a sample of 12 clusters at different redshifts. They find that the scatter varies for different X-ray
observables: from $\sim 23-33\%$ for $T_{\rm X}-M_{\rm WL}$, to $\sim
20-25\%$ for $Y_{\rm X}-M_{\rm WL}$, to $\sim 12-24\%$ for $M_{\rm
gas}-M_{\rm WL}$ relations. The specific values of the scatter depend on
the radius within which the WL masses are measured and typical errors
of the scatter are $\sim 10-15\%$. Based on these results, \citet{okabe2010b} conclude 
that $M_{\rm gas}$ is the best proxy for the total cluster mass. 

One has to ask how scatter as low as $\approx 12\%$ can be obtained
for $M_{\rm gas}-M_{\rm WL,500c}$ relation, given that our results
predict a irreducible scatter of $\approx 18\%$ in $M_{\rm WL,500c}$
with respect to 3D \mfivec\ from triaxiality, correlated, and
uncorrelated LSS (see Fig.~\ref{fig:scatdist}). It is important to note
that measurements of $M_{\rm gas}$ and $M_{\rm WL,500c}$ in the
\citet{okabe2010b} analysis are not independent because gas masses are
measured within \rfivec\ derived from weak lensing mass, $M_{\rm
WL,500c}$. The covariance between $M_{\rm gas}$ and $M_{\rm
WL,500c}$ introduced by this choice can be estimated as follows. Without loss
of generality, let us assume that cumulative gas mass profile can be
locally approximated as $M_{\rm gas}(<r)=M_{\rm gas}(r/\rfivec)^{\alpha_{\rm M}}$
around $r\approx \rfivec$. Given an error in the weak lensing mass $\delta
M_{\rm WL,500c}$ and the corresponding error in radius $\delta
r/\rfivec=(1/3)\delta M_{\rm WL,500c}/M_{\rm WL,500c}$, the fractional
error in the gas mass will be
\begin{displaymath}
\frac{\delta M_{\rm gas}}{M_{\rm gas}}=\alpha_{\rm M}\frac{\delta \rfivec}{\rfivec}=\frac{\alpha_{\rm M}}{3}\frac{\delta M_{\rm WL,500c}}{M_{\rm WL,500c}}\ .
\end{displaymath}
This correlated error in the gas mass from the choice of radial aperture will reduce the apparent scatter of the WL masses at a fixed $M_{\rm gas}$ from the true scatter by the factor of $1-\alpha_{\rm M}/3$. The slope $\alpha_{\rm M}$ at $r\approx \rfivec$ is about $\alpha_{\rm M}\approx 1-1.3$ \citep[measured for clusters in][sample; A. Vikhlinin 2011, private communication]{vikhlinin2006} and the apparent scatter can therefore be reduced by a factor of $\approx 0.55-0.67$. 

Although \citet{okabe2010b} take into account the effect discussed above 
for purely statistical measurement errors of $M_{\rm WL}$,
they do not take into account the intrinsic scatter due to triaxiality, correlated, 
and uncorrelated LSS. Given an intrinsic scatter of $\approx
18\%$ from our analysis and accounting for the effect discussed above, we predict an apparent scatter of $\approx
10-12\%$ -- consistent with the scatter measured by
\citet{okabe2010b} for $M_{\rm WL,500c}-M_{\rm gas}$ relation. We
therefore interpret their measured scatter as simply a manifestation
of the intrinsic scatter of $M_{WL,500c}$ with respect to the 3D masses \mfivec\ and not the true scatter of the
relation between $M_{\rm gas}$ and \mfivec. A simple test of this interpretation is 
to measure scatter of $M_{\rm WL,500c}-M_{\rm gas}$ relation using gas masses measured within $\rfivec$ 
derived from X-ray hydrostatic equilibrium analysis. We predict that the scatter for such relation will increase 
to values close or exceeding its predicted intrinsic value of $\gtrsim 18\%$ (the actual scatter will be somewhat 
larger due to intrinsic scatter between $M_{\rm gas}$ and \mfivec). 

The $M_{\rm WL}-Y_{\rm X}$, and $M_{\rm WL}-T_{\rm
X}$ scaling relations can be interpreted similarly. For example, consider the $M_{\rm WL}-T_{\rm
X}$ scaling relation.  The temperature $T_{\rm X}$ measured in a given radial range can also be sensitive to
errors in $M_{\rm WL}$, if the radial range is defined using radii in units
of $\rfivec$, such as the radial range $0.2-0.5\rfivec$ used by \citet{okabe2010b}.  There is thus also some covariance in the average temperature and 
$M_{\rm WL}$, which will also affect measured scatter in a way qualitatively similar to the scatter in $M_{\rm gas}-M_{\rm WL}$ relation described above. Following a similar assumption that $T_{\rm X}\propto (r/\rfivec)^{\alpha_{\rm T}}$, the measured scatter in the $T_X-M_{\rm WL}$ relation will differ from a true scatter by a factor of $1-\alpha_{\rm T}/3$. Unlike the gas mass profile, the temperature profiles of clusters are generally decreasing with increasing radius outside cluster cores: $\alpha_{\rm T}<0$. Indeed, for the clusters in the \citet{vikhlinin2006} sample, $\alpha_{\rm T}$ ranges from $\alpha_{\rm T}\approx 0$ to $\alpha_{\rm T}\approx -0.2\div -0.4$ (A. Vikhlinin, private communication) for temperatures measured within the radial range $0.2-0.5\rfivec$ used by \citet[][the values of $\alpha_{\rm T}$ measured within $0.15-1\rfivec$ are similar]{okabe2010b}. Potentially, the measured scatter can thus be \textit{increased} by a factor of $\sim 0.10-0.15$. Thus, with an intrinsic scatter in $M_{\rm WL}$ of $18\%$, the measured scatter can be $20-22\%$. The overall effect is small compared to the current errors on the scatter, but may need to be taken into account in the future measurements using much larger samples of clusters. 

For the $M_{\rm WL}-Y_{\rm X}$ scaling relation, we note that the apparent scatter 
will also be affected by covariance between $M_{\rm gas}$ and $M_{\rm WL}$ due to the choice of radial aperture, but that the effect in
this case will be somewhat smaller because the slope of the $M-Y_{\rm X}$
relation is $\approx 0.6$, shallower than almost linear relation
between $M_{\rm WL}-M_{\rm gas}$. In addition, the positive covariance
with $M_{\rm gas}$ can be partially compensated by the negative
covariance with temperature as described above. Therefore, the apparent
scatter of $M_{\rm WL}-Y_{\rm X}$ relation should be closer to
the true intrinsic scatter due to triaxiality and LSS. Indeed, the measured scatter
of $0.200^{+0.066}_{-0.095}$ is consistent with the value of $\approx 0.18$
predicted in our analysis.

The intrinsic scatter of \mwl\ at fixed 3D mass must be taken into account when interpreting the measured scaling relations between different cluster observables and \mwl.  Additionally, there is no contradiction between our results and small scatter measured by \citet{okabe2010b} for the $M_{\rm gas}-M_{\rm WL}$ relation. The scatter measured by these authors can be interpreted as the intrinsic scatter we predict reduced by the covariance between weak lensing and gas mass measured within WL-derived $\rfivec$. Future samples of clusters which will be much larger than the 12 clusters used by \citet{okabe2010b} will test our conclusions and predictions in detail.

\subsection{Implications for Precision Cosmology with Galaxy Clusters and Weak Lensing}\label{sec:cosmology}
Scaling relations between observable
properties of clusters and their total mass are the key component of
cosmological constraints derived from cluster abundances and
clustering. However, the total masses of clusters are notoriously
difficult to measure.  The most common mass estimates, which use
X-ray derived gas and temperature profiles and assumption of
hydrostatic equilibrium (HSE), can be biased low by 5-15\% if non-thermal
pressure support from gas motions or cosmic rays exists in clusters
\citep[see e.g.,][and references therein]{lau2009}. A recent analysis by \citet{mahdavi2008} has indeed indicated that the
HSE X-ray derived masses are biased low by $\approx 10\%$ with respect
to the WL masses, although no such bias has been detected in other
recent studies \citep{zhang2008,vikhlinin2009b,zhang2010}.  Given that the
weak lensing mass estimates are hoped to be used for precise
calibration of cluster mass-observable relations, further independent
studies of systematics and sources of scatter in the WL mass
measurements are critical.

How well the normalization of cluster scaling relations can be
constrained with WL mass measurements depends on the size of the
sample and on the distribution of the measured WL masses with respect
to the 3D cluster mass.  Conversely, if one plans to calibrate an
observable-mass relation to a given accuracy, one needs to know the
scatter and bias to gauge the required cluster sample size.  In this
study we have quantified this distribution using a large cosmological
simulation of a $\Lambda$CDM cosmology. As we discussed in the previous section we 
have measured a scatter of
$\approx30\%$ in the WL mass at fixed 3D mass for parameters which
characterize modern and upcoming WL surveys. This fairly large scatter
implies that samples of at least few dozens of clusters will be
required to constrain the normalization of scaling relations to better
than $5-10\%$. Biases in the WL masses are important as well since
they will directly shift the normalization of cluster scaling
relations.  Biases in WL masses will also complicate attempts to learn
about cluster astrophysics from these scaling relations. WL masses
which are biased low (high) can potentially mask (exacerbate) the
effects of non-thermal pressure support in comparisons with X-ray HSE
mass measurements.

Determining the bias accurately will be more difficult than the scatter.
The current precision to which we can detect intrinsic biases in our
WL estimates of \mfivec\ is $\approx1\%$ at 1-$\sigma$ for the highest
mass halos.  However, we have neglected other systematic and
observational effects which can potentially change the bias in our WL
masses.  Magnification and size bias
\citep[e.g.,][]{schmidt2009a,schmidt2009b} will produce changes in the
relative number of sources in each bin and thus in the properties of
WL mass estimators \citep{schmidt2010,rozo2010}.  Also, we have
assumed perfect knowledge of the source redshifts.  The use of
photometric redshifts or unknown source redshifts can potentially induce biases of
$\approx\pm5-15\%$ in the WL masses as well
\citep[e.g.,][]{mandelbaum2008c,okabe2010}. Contamination of the source galaxies 
by cluster member galaxies can produce $\approx-10\%$ systematic biases as well \citep[e.g.,][]{okabe2010}.
Misidentifying halo centers can also cause small $\approx5\%$ negative biases as well \citep[e.g.,][]{hoekstra2010}.
These and other similar effects will need to be accounted for in order to derive accurate
masses from WL observations.

Note that we have not used any strong lensing information about our
halos.  Using this information would require higher resolution, more
realistic simulations with baryonic physics, full ray tracing of our
halos, and mock galaxies so that realistic multiply imaged galaxies
could be analyzed as is done in observations.  Such an analysis for
three cluster-sized halos have been carried out by
\citet{meneghetti2010a}.  They have found for their three simulated
halos that the inclusion of strong lensing information with the WL
reduced tangential shear profile can reduce the scatter in the
reconstructed cluster masses.  While not all clusters will have strong
lensing features in observations, understanding the statistical
proprieties of strong lensing mass reconstructions in relation to the
population of halos is an important avenue of future research
\citep[see][ for a study of strong lensing cross-sections in this
spirit]{meneghetti2010}.

Finally, our results have implications for follow-up strategies employed with the future cluster samples from large surveys. 
\citet{wu2010} have estimated that follow-up mass observations to check for
systematic errors in self-calibration studies of Dark Energy can
increase the Dark Energy figure of merit (FOM) by up to $40.3-76.4\%$
depending on which clusters as a function of mass and redshift are
selected for follow-up observations. However, such significant
improvements in the FOM can be achieved only if the bias in the
follow-up mass estimates is known to better than $5\%$.  The increase
in the FOM from follow-up observations is less sensitive to the
precise value of the scatter in the follow-up mass estimates, but a
scatter of 40\% can degrade the improvement in the FOM noticeably
compared to 10\%.  Our results put the scatter of WL mass estimates
from ground-based observations near $\approx30\%$, so that some
degradation in the improvement of the FOM compared to their baseline
results is expected.  While in principle we have calibrated the bias
at a level that should not degrade the efficacy of follow-up
observations using WL masses, given its dependence on the exact WL
mass estimation method and other systematic effects not studied here, 
more detailed work is needed in this direction.

\section{Conclusions}\label{sec:conc}

In this paper we have studied the statistical properties of WL mass
estimates obtained by fitting the reduced tangential shear profile
with spherically-symmetric mass model in the thin-lens approximation.  We
have also systematically investigated the sources of scatter and bias in WL
masses as a function of mass and quantified the amount of scatter for
typical ground- and space-based WL observations.  

Importantly, we did not examine in detail the relative merits of using
other spherically-symmetric models or other fitting methods for WL
data. We have found that the scatter and especially bias in WL mass
estimates depends strongly on the specific details of the analysis
like the choice of outer radial fit limit, the choice of halo mass
definition, the degree to which the halo center can be determined
accurately, the choice of fitting metric, and importantly the choice
of WL mass estimation method (i.e. spherically-symmetric model fits,
aperture densitometry, compensated aperture mass estimates, etc.).
Given the large number of choices made when estimating WL masses, it
is difficult to make general statements about the performance of all
WL mass estimators or even explore all the possibilities. We have
chosen to study one common method of WL mass estimation in detail and leave
the investigation of the performance other WL mass estimation methods
to future work.

Our main conclusions are as follows.
\begin{itemize}
\item Weak lensing cluster mass estimates made with spherically-symmetric model fits 
have irreducible scatter from correlated LSS around the clusters in addition to the well-known effects of 
halo triaxiality and uncorrelated LSS.  Specifically, we find that correlated LSS contributes $\approx20\%$ and halo triaxiality 
contributes $\approx80\%$ of the scatter due to matter within $\approx20$ \hmpc\ of the halo center.  
\item For low-mass cluster halos the total intrinsic scatter is dominated by uncorrelated LSS (see Figure~\ref{fig:scatdist} 
and Table~\ref{table:fulllosscatter}).  For the most massive halos, correlated LSS and halo triaxiality are the dominant 
sources of intrinsic scatter for our assumed source redshift $z_{s}=1$.  
\item The contribution of uncorrelated LSS as a function of increasing distance from the cluster for distances $\gtrsim 20$
\hmpc\ is well-described by the formalism of
\citet{hoekstra2001,hoekstra2003}. A similar conclusion was reached by 
\citet{hoekstra2010a}.
\item Weak lensing cluster mass estimates can generally have small, but non-negligible bias of $\approx5-10\%$.  A large portion of this bias is due 
to the fact that the NFW profile commonly assumed in the mass model is not a good
description of the true shear profile at large radii around clusters
from the simulations.  However other physical effects in the
simulations, such as substructure and halo triaxiality, likely
contribute to this bias as well.  We tested an iterative fitting
method, which changes the outer radius of the fit iteratively to the converge to intended virial radius definition (e.g., $r_{500c}$ or $r_{200c}$). Such a method can eliminate most of the bias for \mfivec, but not for masses defined within radii enclosing smaller overdensities. Other methods will therefore need to be explored to eliminate such biases fully. 
\item For current ground-based observations, shape noise is the dominant source of scatter in the weak lensing masses.  
For higher-quality observations with higher source densities, the effects of shape noise become 
comparable or sub-dominant to the effects of the sources of intrinsic scatter: halo triaxiality, correlated LSS, and uncorrelated LSS.  
A similar conclusion was reached by \citet{hoekstra2010a}, but by just comparing shape noise and uncorrelated LSS.
\end{itemize}

The overall implication of our results is that in order to achieve
percent level accuracy in WL mass modeling, the specific details
of both the WL estimation method and observations will matter.  We
will need to include more realistic physical and observational effects
in our simulations.  Additionally, in this work we have used dark
matter only simulations.  While in principle the effects of baryonic
physics on WL mass estimates should be small, the baryonic physics can
affect cluster masses by a few percent \citep{rudd2008}.  Much more
detailed studies will be needed before we can use WL mass estimates to
help study Dark Energy and do precision cosmology.

\acknowledgments We thank Jeremy Tinker for supplying
his halo finder code and Anatoly Klypin for providing us the
simulation used in this study. We also thank Jeremy Tinker, Brant Robertson, 
Masahiro Takada, and especially Alexey Vikhlinin and Eduardo Rozo for their comments and suggestions during this
work. This work was partially supported by the by NASA grant
NAG5-13274, by the Department of Energy grant FNAL08.07, and by Kavli
Institute for Cosmological Physics at the University of Chicago
through grant NSF PHY-0551142 and an endowment from the Kavli
Foundation. This work made extensive use of the NASA Astrophysics Data
System and {\tt arXiv.org} preprint server.

\bibliography{apj-jour,refs}

\end{document}